\pdfoutput=1
\RequirePackage{ifpdf}
\ifpdf 
\documentclass[pdftex]{sigma}
\else
\documentclass{sigma}
\fi

\usepackage{bbm}

\numberwithin{equation}{section}

\newtheorem{Theorem}{Theorem}[section]
\newtheorem{Proposition}[Theorem]{Proposition}
 { \theoremstyle{definition}
\newtheorem{Example}[Theorem]{Example}
\newtheorem{Remark}[Theorem]{Remark} }

\begin{document}
\allowdisplaybreaks

\newcommand{\arXivNumber}{1806.05912}

\renewcommand{\PaperNumber}{087}

\FirstPageHeading

\ShortArticleName{Perturbed $(2n-1)$-Dimensional Kepler Problem}

\ArticleName{Perturbed $\boldsymbol{(2n-1)}$-Dimensional Kepler Problem\\ and the Nilpotent Adjoint Orbits of $\boldsymbol{U(n, n)}$}

\Author{Anatol ODZIJEWICZ}

\AuthorNameForHeading{A.~Odzijewicz}

\Address{Department of Mathematics, University of Bia{\l}ystok,\\
Cio{\l}kowskiego 1M, 15-245 Bia{\l}ystok, Poland}
\Email{\href{mailto:aodzijew@uwb.edu.pl}{aodzijew@uwb.edu.pl}}

\ArticleDates{Received March 11, 2020, in final form September 01, 2020; Published online September 22, 2020}

\Abstract{We study the regularized $(2n-1)$-Kepler problem and other Hamiltonian systems which are related to the nilpotent coadjoint orbits of $U(n,n)$. The Kustaanheimo--Stiefel and Cayley regularization procedures are discussed and their equivalence is shown. Some integrable generalization (perturbation) of $(2n-1)$-Kepler problem is proposed.}


\Classification{53D17; 53D20; 53D22; 70H06}

\section{Introduction}

Kepler problem (not only for historical reasons) is one of the most fundamental subjects of celestial mechanics and quantum mechanics \cite{iwai,toshihiro,malkin,macintosh,mladenov,zwanzi}. Such questions as Moser \cite{moser} and Kustaanheimo--Stiefel \cite{stiefel} regularization procedures as well as the relationship between them~\cite{kummer} are well known for celestial mechanics specialists. Also the questions concerning the quantization of the Kepler system and the MIC-Kepler system, which is its natural generalization, are the subject of many publications, see, e.g., \cite{oh,malkin,meng,mladenov2,mladenov,os,simms}. There are other interesting generalizations of Kepler and MIC-Kepler problems, for example see \cite{bm,iwai3, meng,meng4,meng2,meng3}.

Initially the group $U(2,2)$, being a natural extension of the Poincar\'e group, was recognized as the dynamical group \cite{barut,iwai,iwai2,iwai3, toshihiro,kummer,malkin} for the three-dimensional Kepler and MIC-Kepler problems. Consequently, the group $U(n,n)$ plays the same role for higher dimensional case. Taking this fact into account, in the present paper we study various Hamiltonian systems that have $U(n,n)$ as a dynamical group. They are related to the adjoint nilpotent orbits of $U(n,n)$ and could be interpreted as some natural generalizations of Kepler problem.

In Section \ref{sec:2} we investigate the canonically defined vector bundles over the Grassmannian ${\rm Gr}\big(n, \mathbb{C}^{2n}\big)$ of $n$-dimensional subspaces of twistor space $\mathcal{T}= \big(\mathbb{C}^{2n}, \phi\big)$, where $\phi$ is hermitian form on $\mathbb{C}^{2n}$ with signature $(\underbrace{+\cdots +}_{n}\underbrace{-\cdots -}_{n})$.

In Section \ref{sec:3} we consider the Grassmannian ${\rm Gr}_0 \big(n, \mathbb{C}^{2n}\big)$ of isotropic (with respect to $\phi$) $n$-dimensional subspaces of $\mathbb{C}^{2n}$, which as a manifold is diffeomorphic with the unitary group $U(n)$, and show that $T^*U(n)$ has the structure of $U(n,n)$-Hamiltonian space, see Propositions~\ref{prop:2} and~\ref{prop:3}. In Proposition \ref{prop:3}, we classify the orbits of $U(n,n)$-action on $T^*U(n)$ and specify the one-to-one correspondence of these orbits with such nilpotent adjoint $U(n,n)$-orbits whose elements $\mathfrak{X} $ satisfy $\mathfrak{X}^2 =0$.

In Section \ref{sec:4} we investigate the geometry of the orbit $\mathcal{N}_{1,0}$, which consists of the rank one nilpotent elements of ${\mathfrak u}(n,n)$, see Proposition~\ref{prop:41}. We also discuss the equivalent realizations of the regularized $(2n-1)$-dimensional Kepler problem, see Proposition \ref{prop:nowe}.

In Section \ref{sec:Cayley} we show the equivalence of Cayley and Kustaanheimo--Stiefel regularizations in the context of higher-dimensional Kepler problem, obtaining in this way a natural generalization of the Kustaanheimo--Stiefel transforms for the arbitrary odd dimension.

Finally, in the last Section \ref{sec:5} we consider some integrable generalization of $(2n-1)$-Kepler problem. For this generalized Kepler problem the Hamiltonian, see formula (\ref{65}), depends on the positions and momenta through the coordinates of angular momenta and Runge--Lenz vector. The integrability of this system is proved by the methods developed in \cite{KS}.

\section[Grassmannian ${\rm Gr}\big(n,\mathbb{C}^{2n}\big)$ and related vector bundles]{Grassmannian $\boldsymbol{{\rm Gr}\big(n,\mathbb{C}^{2n}\big)}$ and related vector bundles}\label{sec:2}

In this section we will study some canonically defined bundles over the Grassmannian ${\rm Gr} \big(n, \mathbb{C}^{2n}\big)$ of $n$-dimensional complex vector subspaces of $\mathbb{C}^{2n}$. Let us recall that ${\rm Gr} \big(n, \mathbb{C}^{2n}\big)$ is a $n^2$-dimensional compact complex analytic manifold homogenous with respect to the natural action of ${\rm GL}(2n, \mathbb{C})$.

We begin with defining the following complex analytic bundles over ${\rm Gr}\big( n, \mathbb{C}^{2n}\big)$. Namely, we consider the bundle $\pi_{\mathcal{N}}\colon \mathcal{N} \to {\rm Gr}\big(n, \mathbb{C}^{2n}\big)$ whose fibres consist of nilpotent elements of $\mathfrak{gl}(2n, \mathbb{C})$. The total space of this bundle is defined as
\begin{equation*}
\mathcal{N} := \big\{ (\mathcal{Z}, z) \in \mathfrak{gl}(2n, \mathbb{C}) \times {\rm Gr}\big(n, \mathbb{C}^{2n}\big) \colon \operatorname{Im} (\mathcal{Z}) \subset z \subset \operatorname{Ker} (\mathcal{Z}) \big\}
\end{equation*}
and $\pi_{\mathcal{N}}$ is the projection of $\mathcal{N} $ on the second component of the cartesian product. One easily sees that $\pi_{\mathcal{N}} \colon \mathcal{N} \to {\rm Gr}\big(n,\mathbb{C}^{2n}\big)$ is a complex vector bundle of rank $n^2$. The subset $\operatorname{pr}_1 (\mathcal{N}) \subset \mathfrak{gl}(2n, \mathbb{C})$ consists of such elements $\mathcal{Z}\in \mathfrak{gl}(2n, \mathbb{C})$, which satisfy $\mathcal{Z}^2 = 0$ and have rank $k:= \dim_\mathbb{C} \operatorname{Im} (\mathcal{Z})$, where $1\leq k \leq n$.

Next the bundle $\pi_{\mathcal{P}}\colon \mathcal{P} \to {\rm Gr}\big(n,\mathbb{C}^{2n}\big)$ is a bundle of idempotents, i.e.,
\begin{equation*}
\mathcal{P} := \big\{(p, z) \in \mathfrak{gl}(2n, \mathbb{C}) \times {\rm Gr}\big(n,\mathbb{C}^{2n}\big)\colon p^2 =p,\, \operatorname{Im} (p) = z\big\},
\end{equation*}
where $\pi_{\mathcal{P}}$ is the projection of $\mathcal{P}$ on the second component of cartesian product. We note that $\operatorname{pr}_1(\mathcal{P}) $ consists of such idempotents in $\mathfrak{gl}(2n, \mathbb{C})$ that $\dim_\mathbb{C} (\operatorname{Im} (p)) = n$.

In order to make the structure of $\pi_{\mathcal{P}}\colon \mathcal{P} \to {\rm Gr}\big(n,\mathbb{C}^{2n}\big)$ more transparent, we formulate the following proposition.
\begin{Proposition}\label{pr1}
The bundle $\pi_{\mathcal{P}}\colon \mathcal{P} \!\to \!{\rm Gr}\big(n,\mathbb{C}^{2n}\big)$ is an affine bundle with $\pi_{\mathcal{N}}\colon \mathcal{N} \!\to\! {\rm Gr}\big(n,\mathbb{C}^{2n}\big)$
as the structural vector bundle, i.e., for any $z \in {\rm Gr}\big(n,\mathbb{C}^{2n}\big)$ the vector space $\mathcal{N}_z := \pi^{-1}_\mathcal{N} (z)$ acts in a transitive and free way on the fibre $\mathcal{P}_z := \pi^{-1}_\mathcal{P} (z)$.
\end{Proposition}

\begin{proof}
For $p \in \mathcal{P}_z$ and $\mathcal{Z} \in \mathcal{N}_z$ we have
\begin{equation*}
(p+\mathcal{Z})^2 = p^2 + \mathcal{Z}^2 + p \mathcal{Z} + \mathcal{Z} p = p + \mathcal{Z}
\qquad \text{and}\qquad
(p + \mathcal{Z}) z = pz = z .
\end{equation*}
This shows that $p+\mathcal{Z} \in \mathcal{P}_z$.

For $p, p' \in \mathcal{P}_z$ we have
\begin{equation*}
(p'- p)^2 = p'^2 + p^2 -p'p - pp' = p'+p -p - p' = 0
\end{equation*}
and $\dim_\mathbb{C} \operatorname{Im} (p'-p) \leq n$. Thus, $p'-p =: \mathcal{Z} \in \mathcal{N}_z$. Due to the above facts one has free and transitive action of $\mathcal{N}_z $ on $\mathcal{P}_z$.
\end{proof}

We note that for $p', p \in \mathcal{P}_z$ the following equalities hold
\begin{equation*}
p'p = p \qquad \text{and}\qquad p p' = p'.
\end{equation*}
Subsequently, using the Cartan--Killing form
\begin{equation}\label{CK}
\mathfrak{gl}(2n, \mathbb{C} ) \times \mathfrak{gl}(2n, \mathbb{C}) \ni (\mathcal{Z}_1, \mathcal{Z}_2) \to \operatorname{Tr} (\mathcal{Z}_1 \mathcal{Z}_2) \in \mathbb{C}
\end{equation}
we will identify the dual space $\mathfrak{gl}(2n, \mathbb{C})^* $ with the Lie algebra $\mathfrak{gl}(2n, \mathbb{C})$.

For any $p \in \operatorname{pr}_1(\mathcal{P})$ one has the open subset
\begin{equation*}
\Omega_p := \big\{ z \in {\rm Gr}\big(n,\mathbb{C}^{2n}\big)\colon z \cap (1-p)\mathbb{C}^{2n} = \{0\} \big\}
\end{equation*}
of the Grassmannian. We define a chart $\phi_p \colon \Omega_p \to (1-p)\mathfrak{gl}(2n, \mathbb{C})p \cong \operatorname{Mat}_{n\times n}(\mathbb{C})$ in the following way. The decomposition $z \oplus (1-p) \mathbb{C}^{2n} = \mathbb{C}^{2n}$ defines the projection $q_z $ of $\mathbb{C}^{2n} $ on subspace $z\subset \mathbb{C}^{2n}$. For projections $1-q_z$ and $1-p$ one has $\operatorname{Im} (1-q_z) = \operatorname{Im} (1-p)$. So, according to Proposition~\ref{pr1} there exists $Z\in (1-p)\mathfrak{gl}(2n, \mathbb{C})p$ such that
\begin{equation}\label{12}
Z=(1-p)-(1-q_z)= q_z - p := \phi_p (z).
\end{equation}
The equality \eqref{12} defines the chart $\Omega_p \ni z \mapsto \phi_p(z) =Z$, mentioned above.

In order to find the transition maps $\phi_{p'} \circ \phi_p^{-1} \colon \phi_p(\Omega_{p'} \cap \Omega_{p}) \to \phi_{p'} (\Omega_{p'} \cap \Omega_p)$ between the charts $(\Omega_p, \phi_p )$ and $(\Omega_{p'}, \phi_{p'})$ we observe that for $z\in \Omega_p \cap \Omega_{p'}$ we have
\begin{equation}\label{13}
q_z'q_z = q_z \qquad \text{and} \qquad q_zq_z' = q_z'.
\end{equation}
From \eqref{12} and \eqref{13} we obtain
\begin{gather}
\label{14}
q_z = q_z'q_z= (Z'+p')q_z=p'q_z +Z' p'q_z, \\
\nonumber
q_z = (p+Z)q_z=(p+Z)pq_z= p'(p+Z)pq_z+ (1-p')(p+Z)pq_z \\
\label{kufal}
\hphantom{q_z}{} = p'(p +(1-p)Z)pq_z +(1-p')(p+(1-p)Z)pq_z.
\end{gather}
Expressions \eqref{14} and \eqref{kufal} give two decompositions of $q_z$ on the components from subspaces $ p'\mathfrak{gl} (2n, \mathbb{C})q_z$ and $(1-p')\mathfrak{gl}(2n, \mathbb{C})q_z$, which satisfy $p'\mathfrak{gl} (2n, \mathbb{C})q_z \cap (1-p')\mathfrak{gl}(2n, \mathbb{C})q_z= \{0\}$. So, we have
\begin{gather*}
p'q_z = (a+cZ) pq_z,\qquad
Z'p'q_z = (b+dZ)pq_z,
\end{gather*}
where
\begin{equation*}
a := p'p, \qquad b := (1-p')p , \qquad c:= p'(1-p), \qquad d := (1-p')(1-p).
\end{equation*}
Observing that $p'q_z\colon p\mathbb{C}^{2n} \to p' \mathbb{C}^{2n}$ and $pq_z\colon p\mathbb{C}^{2n} \to p \mathbb{C}^{2n} $ are isomorphisms of the vector subspaces we obtain
\begin{equation*}
Z' = (b+dZ)(a+cZ)^{-1}.
\end{equation*}
Note here that $a+cZ = (p'q_z)(pq_z)^{-1}$. In consequence, the inverse $(a+cZ)^{-1}$ is well defined.
In particular case when $\operatorname{Im} (1-p) = \operatorname{Im} (1-p') $ one has $\Omega_p = \Omega_{p'}$ and, thus
\begin{equation*}
Z' = p - p' +Z .
\end{equation*}

Let us note that one has another canonical complex vector bundles
\begin{equation*}
\mathcal{E} := \big\{ (w, z) \in \mathbb{C}^{2n} \times {\rm Gr} \big(n, \mathbb{C}^{2n} \big) \colon w \in z \big\}
\end{equation*}
and
\begin{equation*}
\mathcal{E}^\bot := \big\{ (\varphi, z) \in \big(\mathbb{C}^{2n}\big)^* \times {\rm Gr} \big(n,\mathbb{C}^{2n}\big) \colon \varphi|_z =0 \big\}
\end{equation*}
over ${\rm Gr}\big(n,\mathbb{C}^{2n}\big)$. The complex linear group ${\rm GL}(2n, \mathbb{C})$ acts on the above bundles in the following way
\begin{gather}\label{ac1}
\Sigma_g (\mathcal{Z}, z) := \big(g\mathcal{Z} g^{-1}, \sigma_g(z)\big),
\\ \label{ac2}
T_g (w,z) := (gw, \sigma_g(z)),
\\ \label{ac3}
T_g^* (\varphi, z) := \big(\varphi \circ g^{-1} , \sigma_g (z)\big),
\end{gather}
where
\begin{equation}\label{sigmag}
\sigma_g (z) := gz,
\end{equation}
for $g\in {\rm GL}(2n, \mathbb{C})$.

The proposition formulated below collects some properties of the above structures which will be useful in the further considerations.
\begin{Proposition}\label{pr4}\quad
\begin{enumerate}\itemsep=0pt
\item[$(i)$] One has the canonical isomorphisms
\begin{equation}\label{iso2}
\mathcal{N} \cong \mathcal{E} \otimes\mathcal{E}^{\bot} \cong T^*{\rm Gr}\big(n,\mathbb{C}^{2n}\big)
\end{equation}
of the ${\rm GL}(2n, \mathbb{C})$-vector bundles.
\item[$(ii)$] The group ${\rm GL}(2n, \mathbb{C})$ acts on $\mathcal{N}$, $\mathcal{E} \otimes\mathcal{E}^{\bot}$ and $T^*{\rm Gr}\big(n,\mathbb{C}^{2n}\big)$ by $\Sigma_g$, $T_g \otimes T_g^*$ and $T^*\sigma_g$, respectively, preserving their vector bundle structures, and isomorphisms from \eqref{iso2} are equivariant with respect to these actions.
\item[$(iii)$] The vector bundle $\mathcal{N}$ $($and thus the vector bundles $\mathcal{E} \otimes\mathcal{E}^{\bot}$ and $T^*{\rm Gr}\big(n,\mathbb{C}^{2n}\big))$ splits into ${\rm GL}(2n, \mathbb{C})$-orbits:
\begin{equation*}
\mathcal{N}^k := \{(\mathcal{Z}, z) \in \mathcal{N} \colon \dim_{\mathbb{C}} \operatorname{Im} \mathcal{Z} = k \},
\end{equation*}
where $k=0,1,\ldots, n$.
\end{enumerate}
\end{Proposition}
\begin{proof}
(i) For proving the isomorphism $\mathcal{N}\cong T^*{\rm Gr}\big(n,\mathbb{C}^{2n}\big)$ we note that
\[
{\rm Gr}\big(n,\mathbb{C}^{2n}\big) \cong {\rm GL}(2n, \mathbb{C})/ {\rm GL}(2n, \mathbb{C})_z,
\]
where the subgroup ${\rm GL}(2n, \mathbb{C})_z$ is the stabilizer of $z\in {\rm Gr}\big(n,\mathbb{C}^{2n}\big)$. Thus, one has
\[T_z {\rm Gr}\big(n,\mathbb{C}^{2n}\big) \cong \mathfrak{gl} (2n, \mathbb{C})/ \mathfrak{gl}(2n, \mathbb{C})_z.\]
Using the isomorphism $\mathfrak{gl}(2n, \mathbb{C})^* \cong \mathfrak{gl}(2n, \mathbb{C})$ defined by the pairing (\ref{CK}), we find that
\[T^*_z {\rm Gr}\big(n,\mathbb{C}^{2n}\big) \cong \left(\mathfrak{gl} (2n, \mathbb{C})/ \mathfrak{gl}(2n, \mathbb{C})_z\right)^* \cong (\mathfrak{gl} (2n, \mathbb{C})_z)^\bot,\]
where $(\mathfrak{gl} (2n, \mathbb{C})_z)^\bot$ is the annihilator of the subspace $\mathfrak{gl} (2n, \mathbb{C})_z \subset \mathfrak{gl} (2n, \mathbb{C})$. Since ${\rm Gr}\big(n, \mathbb{C}^{2n}\big)$ is a ${\rm GL}(2n,\mathbb{C})$-homogenous space, it is enough to prove the above isomorphism for any $z_0\in {\rm Gr}\big(n,\mathbb{C}^{2n}\big)$. So, let us take $z_0 = \left\{\left(\begin{smallmatrix}
\eta \\
0
\end{smallmatrix}\right) \colon \eta \in \mathbb{C}^n\right\}$ for which{\samepage
\[\mathfrak{gl}(2n, \mathbb{C})_{z_0} = \left\{ \left(\begin{matrix}
A & B \\
0 & D
\end{matrix}\right) \in \operatorname{Mat}_{2n \times 2n}(\mathbb{C})\colon A, B, D \in \operatorname{Mat}_{n\times n} (\mathbb{C}) \right\}.\]
One easily sees that for such $z_0$ we have $(\mathfrak{gl}(2n, \mathbb{C})_{z_0})^\bot \cong \mathcal{N}_{z_0}$.}

In order to prove that $\mathcal{E}_z \otimes\mathcal{E}_z^{\bot} \cong \mathcal{N}_z$ we observe that $\mathcal{Z} \in \mathcal{E}_z \otimes\mathcal{E}_z^{\bot} \subset \mathfrak{gl}(2n, \mathbb{C})$ satisfies $\operatorname{Im} \mathcal{Z} \subset z \subset \operatorname{Ker} \mathcal{Z}$ and $\dim_{\mathbb{C}} \mathcal{E}_z \otimes\mathcal{E}_z^{\bot} = \dim_{\mathbb{C}} \mathcal{N}_z$.

(ii) The ${\rm GL}(2n, \mathbb{C})$-equivariance of the above bundles isomorphisms one easily see from \mbox{(\ref{ac1})--(\ref{sigmag})}.

(iii) By straightforward verification.
\end{proof}

From Proposition \ref{pr4} we conclude:
\begin{Remark}\label{rem1}\quad
\begin{enumerate}\itemsep=0pt
\item[(i)] The orbit $\mathcal{N}^0$ is the zero section of $\mathcal{N} \to {\rm Gr}\big(n,\mathbb{C}^{2n}\big)$ so, one can identify it with ${\rm Gr}\big(n,\mathbb{C}^{2n}\big)$.
\item[(ii)] The orbit $\mathcal{N}^n $ is an open-dense subset of $\mathcal{N}$.
\end{enumerate}
\end{Remark}

We mention here that $\mathcal{N}^k$ is the total space of the following ${\rm GL}(2n, \mathbb{C})$-homogeneous bundles:
\begin{equation*}
\hspace*{7.5mm}\unitlength=5mm\begin{picture}(11,4.6)
 \put(8,4){\makebox(0,0){$\mathcal{N}^k$}}

 \put(15,0){\makebox(0,0){${\rm Gr}\big(k, \mathbb{C}^{2n}\big)$,}}
 \put(8,0){\makebox(0,0){${\rm Gr}\big(n,\mathbb{C}^{2n}\big)$}}
 \put(1,0){\makebox(0,0){${\rm Gr}\big(n-k, \mathbb{C}^{2n}\big)$}}

 \put(8,3){\vector(0,-1){2}}
 \put(7,3){\vector(-2,-1){4}}
 \put(9,3){\vector(2,-1){4}}

 \put(2,2){\makebox(0,0){$\pi_{\rm ker}$}}
 \put(9,2){\makebox(0,0){$\pi$}}
\put(13,2){\makebox(0,0){$ \pi_{\rm im}$}}
 \end{picture}\end{equation*}
		
where the bundle projections are defined by
\begin{gather*}
\pi_{\rm im} (\mathcal{Z}, z) = \operatorname{Im} (\mathcal{Z}), \qquad
\pi (\mathcal{Z}, z) = z,\qquad
\pi_{\rm ker} (\mathcal{Z}, z) = \operatorname{Ker} (\mathcal{Z}) .
\end{gather*}

\begin{Remark}\label{rem2}In the Penrose twistor theory, see, e.g., \cite{penrose}, which concerns the case $n=2$, the submanifolds $\pi \big(\pi^{-1}_{\rm im} (z)\big)$ and $\pi \big(\pi^{-1}_{\rm ker} (z)\big)$ are called the $\alpha$-planes and $\beta$-planes, respectively.
\end{Remark}

\section[$T^*U(n)$ as a Hamiltonian $U(n,n)$-space]{$\boldsymbol{T^*U(n)}$ as a Hamiltonian $\boldsymbol{U(n,n)}$-space}\label{sec:3}

Now we will describe some real versions of the structures described in the previous section and their relation to the structure of the cotangent bundle $T^*U(n)$ as a $U(n,n)$-Hamiltonian space. For this reason we fix a scalar product
\begin{equation}\label{sproduct}
\langle v, w \rangle := v^+ \phi w
\end{equation}
of $v,w \in \mathbb{C}^{2n}$, defined by a hermitian matrix $\phi = \phi^+ \in \operatorname{Mat}_{2n\times 2n} (\mathbb{C})$ which has signature $(\underbrace{+ \cdots +}_{n}\underbrace{-\cdots - }_{n} )$ and satisfies $\phi^2 = \mathbbm{1}_{2n}$. By ``$+$'' in (\ref{sproduct}) and everywhere below we will denote the hermitian transposition of a matrix. Hence we define the group $U(n,n)$ and Lie algebra ${\mathfrak u}(n,n) $ of $U(n,n)$ by
\begin{equation*}
g^+ \phi g = \phi
\end{equation*}
and by
\begin{equation*}
\mathfrak{X}^+ \phi + \phi \mathfrak{X} = 0,
\end{equation*}
respectively, where by definition $g \in U(n,n) $ and $\mathfrak{X} \in {\mathfrak u}(n,n)$. Since for $n=2$ the vector space~$\mathbb{C}^{2n}$ provided with scalar product~(\ref{sproduct}) is known as twistor space~\cite{penrose}, in the subsequent we will use the same terminology for an arbitrary dimension.

Using scalar product \eqref{sproduct} we also define on $\mathfrak{gl}(2n, \mathbb{C})$, $ {\rm Gr}\big(n,\mathbb{C}^{2n}\big)$ and $\mathcal{N}$, respectively, the following involutions
\begin{gather}\label{in1}
I(\mathcal{Z}) := - \phi \mathcal{Z}^+ \phi ,\\
\label{in2}
\bot (z) := z^\bot,\\
\label{in3}
\tilde{I} (\mathcal{Z}, z) := \big(I(\mathcal{Z}), z^\bot \big) ,
\end{gather}
where $z^\bot \subset \mathbb{C}^{2n} $ is the orthogonal complement of $z \in {\rm Gr}(n, \mathbb{C})$ with respect to \eqref{sproduct} and $\mathcal{Z} \in \mathfrak{gl}(2n, \mathbb{C})$. Let us note that~\eqref{in1} is an anti-linear map of $\mathfrak{gl}(2n, \mathbb{C})$ and \eqref{in3} is a fibre-wise anti-linear map of the bundle $\pi_{\mathcal{N}}\colon \mathcal{N} \to {\rm Gr}\big(n,\mathbb{C}^{2n}\big)$. Hence, taking into account the equivalent equalities
\begin{equation}\label{37}
\operatorname{Im}(I(\mathcal{Z})) = (\operatorname{Ker}(\mathcal{Z}))^\bot \qquad \text{and} \qquad \operatorname{Ker}(I(\mathcal{Z})) = (\operatorname{Im}(\mathcal{Z}))^\bot
\end{equation}
we obtain the anti-holomorphic bundle isomorphism
 \begin{equation*}\unitlength=5mm 
 \hspace*{9.5mm} \begin{picture}(11,4.6)
 \put(1,4){\makebox(0,0){$\mathcal{N}$}}
 \put(8,4){\makebox(0,0){$\mathcal{N}$}}
 \put(0,0){\makebox(0,0){${\rm Gr}\big(n,\mathbb{C}^{2n}\big)$}}
 \put(9,0){\makebox(0,0){${\rm Gr}\big(n,\mathbb{C}^{2n}\big)$,}}
 \put(0.7,3){\vector(0,-1){2}}
 \put(8.2,3){\vector(0,-1){2}}
 \put(2.5,4){\vector(1,0){3.7}}
 \put(2.7,0){\vector(1,0){3.7}}

 \put(4.5,4.5){\makebox(0,0){$\tilde{I}$}}
 \put(4.5,0.5){\makebox(0,0){$ \bot $}}
 \end{picture}\end{equation*}
which restricts to the isomorphism
\begin{equation*}\unitlength=5mm
\hspace*{9.5mm}\begin{picture}(11,4.6)
 \put(1,4){\makebox(0,0){$\mathcal{N}^k$}}
 \put(8,4){\makebox(0,0){$\mathcal{N}^{k}$}}
 \put(0,0){\makebox(0,0){${\rm Gr}\big(n,\mathbb{C}^{2n}\big)$}}
 \put(9,0){\makebox(0,0){${\rm Gr}\big(n,\mathbb{C}^{2n}\big)$}}
 \put(0.7,3){\vector(0,-1){2}}
 \put(8.2,3){\vector(0,-1){2}}
 \put(2.5,4){\vector(1,0){3.7}}
 \put(2.7,0){\vector(1,0){3.7}}

 \put(4.5,4.5){\makebox(0,0){$\tilde{I}$}}
 \put(4.5,0.5){\makebox(0,0){$ \bot $}}
 \end{picture}\end{equation*}
of the bundle $\mathcal{N}^k \to {\rm Gr}\big(n,\mathbb{C}^{2n}\big)$. Note here that the bundle $\mathcal{N}^k \to {\rm Gr}\big(n,\mathbb{C}^{2n}\big)$ is not a vector bundle. All the above isomorphisms are equivariant with respect to the actions of $U(n,n)\subset {\rm GL}(2n, \mathbb{C})$ defined in \eqref{ac1} and \eqref{sigmag}.

By $\pi_{\mathcal{N}_0} \colon \mathcal{N}_0 \to {\rm Gr}_0\big(n,\mathbb{C}^{2n}\big)$ we denote the vector bundle over the Grassmannian ${\rm Gr}_0\big(n,\mathbb{C}^{2n}\big)$ of complex $n$-dimensional vector subspaces, which are isotropic with respect to the scalar product~\eqref{sproduct}, i.e., $z\in {\rm Gr}_0 \big(n,\mathbb{C}^{2n}\big)$ if and only if $z = z^\bot $. The total space $\mathcal{N}_0$ of this bundle is defined as the subset $\mathcal{N}_0 \subset \mathcal{N}$ of fixed points of the involution $\tilde{I} \colon \mathcal{N} \to \mathcal{N}$ defined in~\eqref{in3}. Let us note here that $\dim_\mathbb{R} {\rm Gr}_0 \big(n,\mathbb{C}^{2n}\big) = n^2$.

We define the map of the vector bundle $\mathcal{N}_0$ into the Lie algebra ${\mathfrak u}(n,n)$ by
\begin{equation*}
\operatorname{pr}_1 (\mathfrak{X}, z) := \mathfrak{X}.
\end{equation*}
The set of values of this map is determined in the following way.

\begin{Proposition}\label{prop:1}
An element $\mathfrak{X} \in {\mathfrak u}(n,n)$ belongs to $\operatorname{pr}_1 (\mathcal{N}_0)$ if and only if $\mathfrak{X}^2 =0$.
\end{Proposition}

\begin{proof}If $\mathfrak{X} \in {\mathfrak u}(n,n) $ satisfies $\mathfrak{X}^2 =0$ then because of $I(\mathfrak{X}) = \mathfrak{X}$ and (\ref{37}) we find that
\[
\operatorname{Im}(\mathfrak{X}) \subset \operatorname{Ker} (\mathfrak{X}) = (\operatorname{Im} (\mathfrak{X}))^\bot .
\]
From the above and nonsingularity of the scalar product (\ref{sproduct}) we obtain
\[
k:= \dim_\mathbb{C} \operatorname{Im} (\mathfrak{X}) \leq \dim_\mathbb{C} (\operatorname{Im} (\mathfrak{X}))^\bot = 2n-k.
\]
So, $0\leq k \leq n$ and thus, there exists $z \in {\rm Gr}_0 \big(n,\mathbb{C}^{2n}\big)$ such that
\begin{equation}\label{eq:311}
\operatorname{Im} (\mathfrak{X}) \subset z \subset \operatorname{Ker} (\mathfrak{X}),
\end{equation}
i.e., $\mathfrak{X} \in \operatorname{pr}_1(\mathcal{N}_0)$. Note that $\operatorname{Im} (\mathfrak{X})$ as an isotropic subspace of $\big(\mathbb{C}^{2n}, \langle \cdot, \cdot \rangle \big)$ could be extended to maximal isotropic subspace~$z$, which has dimension~$n$ and is contained in $(\operatorname{Im}(\mathfrak{X}))^\bot $.

By definition of $\mathcal{N}_0$ any element $\mathfrak{X} \in \operatorname{pr}_1 (\mathcal{N}_0)$ satisfies (\ref{eq:311}), so one has $\mathfrak{X}^2 =0$.
\end{proof}

Next, taking the decomposition $\mathbb{C}^{2n} = \mathbb{C}^n \oplus \mathbb{C}^n $, we will choose the hermitian matrix from the definition (\ref{sproduct}) in the following diagonal block form
\begin{equation}\label{eq:312n}
\phi_d = \left(\begin{matrix}
E & 0\\
0 & -E
\end{matrix}\right),
\end{equation}
 where $E$ and $0$ are unit and zero $n\times n$-matrices. Hence, we obtain
\begin{equation}\label{eq:312}
\langle v, v\rangle = \eta^+ \eta - \xi^+ \xi ,
\end{equation}
for $v = \left(\begin{smallmatrix}
\eta \\
\xi
\end{smallmatrix}\right) \in \mathbb{C}^n \oplus \mathbb{C}^n$.

Now, let us take a set $\left\{v_1 = \left(\begin{smallmatrix}
\eta_1 \\
\xi_1
\end{smallmatrix}\right), \ldots , v_n =\left(\begin{smallmatrix}
\eta_n \\
\xi_n
\end{smallmatrix}\right)\right\} \subset \mathbb{C}^{2n} $ of linearly independent vectors which span $z \in {\rm Gr}_0 \big(n,\mathbb{C}^{2n}\big)$. Since $\langle v_k , v_l \rangle =0$ it follows that
\begin{equation}\label{eq:313}
\eta_k^+\eta_l = \xi_k^+ \xi_l
\end{equation}
for $k,l = 1,\ldots, n$. From (\ref{eq:313}) we see that there exists such $Z \in U(n)$ that $\eta_k = Z\xi_k $ for $k=1,\ldots, n$. So, vectors $\xi_1, \ldots, \xi_n $ form a basis in $\mathbb{C}^n$.

The above considerations show that there is a natural diffeomorphism $U(n) \cong {\rm Gr}_0 \big(n,\mathbb{C}^{2n}\big)$ between the unitary group $U(n)$ and the Grassmannian ${\rm Gr}_0\big(n,\mathbb{C}^{2n}\big)$ of $n$-dimensional isotropic subspaces of $\big(\mathbb{C}^{2n}, \langle \cdot, \cdot \rangle \big)$, which is defined in the following way
\begin{equation}\label{eq:314}
I_0 \colon \ U(n) \ni Z \mapsto z :=\left\{\left(\begin{matrix}
Z\xi \\
\xi
\end{matrix}\right) \colon \xi \in \mathbb{C}^n\right\} \in {\rm Gr}_0 \big(n , \mathbb{C}^{2n}\big) .
\end{equation}

One easily sees that for $\phi_d $ the block matrix elements $A,B,C,D \in \operatorname{Mat}_{n\times n} (\mathbb{C}) $ of $g = \left(\begin{smallmatrix}
A & B\\
C & D
\end{smallmatrix}\right) \in U(n,n)$ satisfy
\begin{equation}\label{eq:315}
A^+A = E+ C^+C, \qquad D^+D = E+ B^+B \qquad\text{and} \qquad D^+C = B^+A .
\end{equation}
From (\ref{eq:314}) one finds that $U(n,n)$ acts on $U(n)$ as follows
\begin{equation}\label{eq:316}
Z' = \sigma_g (Z) = (AZ+B)(CZ+D)^{-1}.
\end{equation}
Subsequently we will need the explicit description of the stabilizer $U(n,n)_E := \{ g \in U(n,n)\colon$ $\sigma_g (E) = E \} $ of the group unit $E\in U(n)$. Simple considerations shows that $g = \left(\begin{smallmatrix}
A & B\\
C & D
\end{smallmatrix}\right) \in U(n,n)_E$ if and only if
\begin{gather}
A = \tfrac{1}{2} \big(\big(F^+\big)^{-1} + F \big) + H,\qquad
D = \tfrac{1}{2} \big(\big(F^+\big)^{-1} + F \big) - H,\nonumber\\
B = \tfrac{1}{2} \big( F- \big(F^+\big)^{-1} \big) - H,\qquad
C = \tfrac{1}{2} \big( F- \big(F^+\big)^{-1} \big) + H,\label{eq:317}
\end{gather}
where $F\in {\rm GL}(n, \mathbb{C})$ and $H \in \operatorname{Mat}_{n\times n}(\mathbb{C})$ satisfy the condition
\begin{equation*}
HF^+ + FH^+ =0.
\end{equation*}

Let us take a smooth curve $]{-}\epsilon , \epsilon [{} \ni t \mapsto Z(t) \in U(n)$ through the element $Z= Z(0)$. By $\dot{Z}:=\frac{{\rm d}}{{\rm d}t} Z(t) |_{t=0} \in T_Z U(n)$ we denote the vector tangent to $Z(t)$ at $Z$ and by $\tau := Z^{-1} \dot{Z} \in T_E U(n)\cong {\mathfrak u}(n) \cong {\rm i} H(n)$, where by $H(n) $, we denote the real vector space of $(n\times n)$-hermitian matrices. Using the above notation, from~\eqref{eq:316} and~(\ref{eq:315}) we obtain
\begin{gather}
\tau ' = Z'^{-1} \dot{Z}' = \big((AZ+B)(CZ+D)^{-1}\big)^+\big(A\dot{Z}(CZ+D)^{-1} \nonumber\\
\hphantom{\tau ' =}{} -(AZ+B)(CZ+D)^{-1}C\dot{Z}(CZ+D)^{-1}\big)\nonumber\\
\hphantom{\tau '}{}= \big((AZ+B)(CZ+D)^{-1}\big)^+A\dot{Z}(CZ+D)^{-1} - C\dot{Z}(CZ+D)^{-1} \nonumber\\
\hphantom{\tau '}{}=\big((CZ+D)^{-1}\big)^+(Z^+\big(A^+A-C^+C\big)\dot{Z}+ \big(B^+A-D^+C\big)\dot{Z})(CZ+D)^{-1} \nonumber\\
\hphantom{\tau '}{}=\big((CZ+D)^{-1}\big)^+Z^+\dot{Z} (CZ+D)^{-1} = \big((CZ+D)^{-1}\big)^+ \tau (CZ+D)^{-1}.\label{eq:318}
\end{gather}
Since $\dot{Z} = Z\tau$, we have the isomorphism of vector bundles $TU(n) \cong U(n) \times {\rm i}H(n)$. It follows from (\ref{eq:318}) that the covector $\rho \in T_E^*U(n) \cong {\rm i}H(n)$ transforms in the following way
\begin{equation*}
\rho' = (CZ+D) \rho (CZ+D)^+ ,
\end{equation*}
where one identifies the Lie algebra $({\rm i}H(n), [\cdot, \cdot ])$ of $U(n)$ with its dual $({\rm i}H(n))^*$ by the Cartan--Killing form.

The elements of Lie algebra ${\mathfrak u}(n,n)$ in the diagonal realization (\ref{eq:312}) of $\phi$ are given by matrices
\begin{gather*}
\mathfrak{X} = \left(\begin{matrix}
\alpha & \beta \\
\beta^+ & \delta
\end{matrix}\right) ,
\end{gather*}
where $\beta \in \operatorname{Mat}_{n\times n} (\mathbb{C})$ and $\alpha, \delta \in {\rm i}H(n) $.

\begin{Proposition}\label{prop:2}\quad
\begin{enumerate}\itemsep=0pt
\item[$(i)$] The map ${\bf I}_0 \colon T^* U(n) \cong U(n) \times {\rm i} H(n) \to \mathcal{N}_0$ defined by
\begin{equation}\label{eq:321}
{\bf I}_0 (Z, \rho ) := \left(\left(\begin{matrix}
-Z\rho Z^+ & Z\rho \\
(Z\rho)^+ & \rho
\end{matrix}\right) , \left\{\left(\begin{matrix}
Z\xi\\
\xi
\end{matrix}\right) \colon \xi \in \mathbb{C}^n \right\}\right) \in \mathcal{N}_0
\end{equation}
is a $U(n,n)$-equivariant
\begin{equation*}
{\bf I}_0\circ \Lambda_g= \Sigma_g \circ {\bf I}_0
\end{equation*}
isomorphism of the vector bundles. The action $\Sigma_g \colon \mathcal{N}_0 \to \mathcal{N}_0 $, $g\in U(n,n) $, is a restriction to~$U(n,n)$ and $\mathcal{N}_0 \subset \mathcal{N} $ of the action defined in~\eqref{ac1}. The action $\Lambda_g \colon U(n) \times {\rm i} H(n) \to U(n) \times {\rm i}H(n) $ is defined by
\begin{equation}\label{eq:323}
\Lambda_g (Z, \delta ) = \big((AZ+B)(CZ+D)^{-1}, (CZ+D)\delta (CZ+D)^+\big),
\end{equation}
where $g = \left(\begin{smallmatrix}
A & B\\
C & D
\end{smallmatrix}\right) $.
\item[$(ii)$] The canonical one-form $\gamma_0 $ on $T^*U(n)\cong U(n) \times {\rm i}H(n)$ written in the coordinates $(Z, \delta) \in U(n)\times {\rm i}H(n)$ assumes the form
\begin{equation}\label{eq:324}
\gamma_0 = {\rm i}\operatorname{Tr} (\rho Z^+ {\rm d}Z)
\end{equation}
and it is invariant with respect to the action \eqref{eq:323}.

\item[$(iii)$] The map ${\bf J}_0\colon T^*U(n) \to {\mathfrak u}(n,n)$ defined by
\begin{equation}\label{eq:326n}
{\bf J}_0 (Z, \rho ):= (\operatorname{pr}_1 \circ {\bf I}_0 )(Z, \rho)= \left(\begin{matrix}
-Z\rho Z^+ & Z\rho \\
(Z\rho )^+ & \rho \end{matrix}\right)
\end{equation}
is the momentum map for symplectic form ${\rm d}\gamma_0$, i.e., it is a $U(n,n)$-equivariant
\begin{equation}\label{eq27}
\operatorname{Ad}_g \circ {\bf J}_0 = {\bf J}_0 \circ \Lambda_g
\end{equation}
Poisson map of symplectic manifold $(T^* U(n), {\rm d}\gamma_0)$ into Lie--Poisson space $({\mathfrak u}(n,n) \cong {\mathfrak u}(n,n)^*, \{ \cdot , \cdot \}_{\text{\rm L-P}})$, where
\begin{gather}
\{f, g\}_{{\text{\rm L-P}}} (\alpha, \delta , \beta , \beta^+ ) = \operatorname{Tr} \Bigg(\alpha\left(\left[\frac{\partial f}{\partial \alpha }, \frac{\partial g}{\partial \beta}\right] +\frac{\partial f}{\partial \beta }\frac{\partial g}{\partial \beta^+ } - \frac{\partial g}{\partial \beta }\frac{\partial f}{\partial \beta^+}\right) \nonumber\\
\hphantom{\{f, g\}_{{\text{\rm L-P}}} (\alpha, \delta , \beta , \beta^+ ) =}{} + \beta\left(\frac{\partial f}{\partial \beta^+ }\frac{\partial g}{\partial \alpha } + \frac{\partial f}{\partial \delta }\frac{\partial g}{\partial \beta^+ } - \frac{\partial g}{\partial \beta^+ }\frac{\partial f}{\partial \alpha }- \frac{\partial g}{\partial \delta }\frac{\partial f}{\partial \beta^+ }\right) \nonumber\\
\hphantom{\{f, g\}_{{\text{\rm L-P}}} (\alpha, \delta , \beta , \beta^+ ) =}{} + \beta^+ \left(\frac{\partial f}{\partial \alpha }\frac{\partial g}{\partial \beta } + \frac{\partial f}{\partial \beta }\frac{\partial g}{\partial \delta } - \frac{\partial g}{\partial \alpha }\frac{\partial f}{\partial \beta }- \frac{\partial g}{\partial \beta }\frac{\partial f}{\partial \delta}\right)\nonumber\\
\hphantom{\{f, g\}_{{\text{\rm L-P}}} (\alpha, \delta , \beta , \beta^+ ) =}{}
+ \delta\left(\left[\frac{\partial f}{\partial \delta }, \frac{\partial g}{\partial \delta }\right] + \frac{\partial f}{\partial \beta^+ }\frac{\partial g}{\partial \beta }- \frac{\partial g}{\partial \beta^+ }\frac{\partial f}{\partial \beta }\right)\Bigg)\label{eq:325}
\end{gather}
for $f,g \in C^\infty ({\mathfrak u}(n,n), \mathbb{R})$.
\end{enumerate}
\end{Proposition}

\begin{proof}(i) From the definition of $\mathcal{N}_0$ it follows that $(\mathfrak{X}, z ) \in \mathcal{N}_0$ if and only if it satisfies~(\ref{eq:311}). Thus, using $U(n) \cong {\rm Gr}_0 \big(n,\mathbb{C}^{2n}\big)$ and~(\ref{eq:321}) we find that $\beta=Z \delta$ and $\alpha = -Z\delta Z^+$ for $\mathfrak{X}\in \operatorname{pr}_1 (\mathcal{N}_0)$. The above shows that ${\bf I}_0\colon U(n) \times {\rm i}H(n) \to \mathcal{N}_0 $ is an isomorphism of vector bundles.

One proves the equivariance property (\ref{eq:323}) by straightforward verification.

(ii) One obtains (\ref{eq:324}) directly from the definition of canonical form $\gamma_0$ on $T^* U(n)$ and from the isomorphism $T^*U(n) \cong U(n)\times {\rm i} H(n)$.

(iii) The equivariance property \eqref{eq27} and formula (\ref{eq:325}) for Lie--Poisson bracket follow by straightforward verification.
\end{proof}

Now, we will describe the relation between the $\operatorname{Ad}(U(n,n))$-orbits in $\operatorname{pr}_1(\mathcal{N}_0)$ and $\Lambda (U(n,n))$-orbits in $T^*U(n)$. We present the most important facts in the following proposition.

\begin{Proposition}\label{prop:3}\quad
\begin{enumerate}\itemsep=0pt
\item[$(i)$] Any $\Lambda (U(n,n))$-orbit $\mathcal{O}_{k,l} $ in $T^*U(n)= U(n) \times {\rm i} H(n)$ is uniquely generated from the element $(E, \rho_{k,l})\in U(n) \times {\rm i} H(n)$, where
\begin{equation}\label{eq:326}
\rho_{k,l} := {\rm i} \operatorname{diag} (\underbrace{1, \ldots , 1}_{k}\underbrace{-1, \ldots , -1}_{l}\underbrace{0, \ldots, 0}_{n-k-l} )
\end{equation}
and has structure of a trivial bundle $\mathcal{O}_{k,l} \to U(n)$ over $U(n)$, i.e., $\mathcal{O}_{k,l}\cong U(n)\times \Delta_{k,l}$, where $\Delta_{k,l} := \{ F \rho_{k,l} F^+ \colon F \in {\rm GL}(n, \mathbb{C})\}$.
\item[$(ii)$] The momentum map \eqref{eq:326n} gives one-to-one correspondence $\mathcal{O}_{k,l} \leftrightarrow {\bf J}_0 (\mathcal{O}_{k,l}) = \mathcal{N}_{k,l} \subset \operatorname{pr}_1 (\mathcal{N}_0) = \big\{\mathfrak{X} \in {\mathfrak u}(n,n) \colon \mathfrak{X}^2 =0 \big\}$ between $\Lambda (U(n,n))$-orbits in $T^*U(n)$ and $\operatorname{Ad}(U(n,n))$-orbits in $\operatorname{pr}_1 (\mathcal{N}_0)$, where $\mathcal{N}_{k,l}= \{ \operatorname{Ad}_g{\bf I}_0(E, \rho_{k,l}) \colon g \in U(n,n)\}$.
\end{enumerate}
\end{Proposition}
\begin{proof}(i) Since the action of $U(n,n)$ on $U(n)$ is transitive, one can identify any $\Lambda (U(n,n))$-orbit~$\mathcal{O} $ in $T^*U(n) \cong U(n)\times {\rm i}H(n)$ with $U(n) \times \Delta$, where $\Delta $ is an orbit of $U(n,n)_E$ in $T^*_E U(n) \cong {\rm i}H(n)$. The action of $g\in U(n,n)_E$, which is defined in~(\ref{eq:317}), on $(E, \rho ) \in \{E \} \times {\rm i}H(n)$ is given by
\begin{equation}\label{eq:330}
\Lambda_g (E, \rho ) = \big(E, F\rho F^+\big),
\end{equation}
 where $F \in {\rm GL}(n, \mathbb{C})$. From (\ref{eq:330}) and Sylvester signature theorem, see~\cite{lang}, follows that $\Delta $ has form $\Delta_{kl} := \{ F\rho_{k,l} F^+ \colon F \in {\rm GL}(n, \mathbb{C})\}$, where $\rho_{k,l} $ is defined in~(\ref{eq:326}).

(ii) From Proposition \ref{prop:1} and point (i) of Proposition \ref{prop:2} it follows that any $\operatorname{Ad}(U(n,n))$-orbit in $\operatorname{pr}_1 (\mathcal{N}_0)$ has form ${\bf J}_0 (\mathcal{O}_{k,l})$. Since for $g\in U(n,n)_E$ we have
\begin{equation*}
\operatorname{Ad}_g({\bf J}_0 (E, \rho)) = {\bf J}_0 (\Lambda_g (E, \rho)) = {\bf J}_0 \big(E, F\rho F^+\big),
\end{equation*}
the momentum map ${\bf J}_0 \colon T^*U(n) \to \operatorname{pr}_1 (\mathcal{N}_0)$ maps $\mathcal{O}_{k,l}$ on the one $\operatorname{Ad}(U(n,n))$-orbit $\mathcal{N}_{k,l}\subset \operatorname{pr}_1 (\mathcal{N}_0)$ only.
\end{proof}

As it follows from general theory, the $\operatorname{Ad}(U(n,n))$-orbit $\mathcal{N}_{k,l}$ is a homogenous symplectic manifold with the symplectic form $\omega_{kl}$, obtained in a canonical way by Kirillov construction, see \cite{kirillov}. From point (ii) of Proposition~\ref{prop:3} we have ${\bf J}_0^{-1} (\mathcal{N}_{k,l}) = \mathcal{O}_{k,l}$.
Hence, one can obtain $(\mathcal{N}_{k,l}, \omega_{kl})$ reducing standard symplectic form ${\rm d}\gamma_0$ on $T^*U(n)$ to the orbit $\mathcal{O}_{k,l}$. Let us note here that fibres ${\bf J}_0^{-1} (\mathfrak{X})$, $\mathfrak{X} \in \mathcal{N}_{k,l}$, are degeneracy submanifolds for the $2$-form ${\rm d}\gamma_0 |_{\mathcal{O}_{k,l}}$, so, $\mathcal{N}_{k,l}= \mathcal{O}_{k,l}/_\sim $ and $\omega_{kl} = {\rm d}\gamma_0|_{\mathcal{O}_{k,l}}/_\sim $, where ``$\sim$'' is an equivalence relation on $\mathcal{O}_{k,l}$ defined by the submersion ${\bf J}_0\colon \mathcal{O}_{k,l} \to \mathcal{N}_{k,l}$.

Ending this section, we mention that in the case when $k+l=n$ one has $\mathcal{N}_{k,l} \cong \mathcal{O}_{k,l}$ and the orbits $ \mathcal{O}_{k,l}$ are open subsets of the cotangent bundle~$T^*U(n)$. For symplectic forms $\omega_{kl}$ we have $\omega_{kl}= {\rm d}\gamma_0$.

For $k=l=0$ the orbit $\mathcal{O}_{00} \cong U(n)$ is the zero section of $T^*U(n)$ and ${\bf J}_0(\mathcal{O}_{00})=\mathcal{N}_{00}= \{0\}$.

\section[Regularized $(2n-1)$-dimensional Kepler problem]{Regularized $\boldsymbol{(2n-1)}$-dimensional Kepler problem}\label{sec:4}

In this section we will describe in detail the various Hamiltonian systems having $U(n,n)$ as their dynamical group. As we will show in the next section, these systems give the equivalent description of the regularized $(2n-1)$-dimensional Kepler system.

Let us begin by defining $U(n,n)$-invariant differential one-form
\begin{equation}\label{41}
\gamma_{+-} := {\rm i} \big(\eta^+ {\rm d}\eta - \xi^+ {\rm d}\xi \big)
\end{equation}
on $\mathbb{C}^{2n} = \mathbb{C}^n \oplus \mathbb{C}^n $.
The Poisson bracket $\{f, g\}_{+-}$ and momentum map ${\bf J}_{+-}\colon \mathbb{C}^{2n} \to {\mathfrak u}(n,n)$ corresponding to the symplectic form ${\rm d}\gamma_{+-}$ are given by
\begin{equation*}
\{f, g\}_{+-} := {\rm i}\left(\frac{\partial f}{\partial \eta^+ }\frac{\partial g}{\partial \eta} - \frac{\partial g}{\partial \eta^+ }\frac{\partial f}{\partial \eta}
- \left(\frac{\partial f}{\partial \xi^+ }\frac{\partial g}{\partial \xi}- \frac{\partial g}{\partial \xi^+ }\frac{\partial f}{\partial \xi}\right)\right)
\end{equation*}
and by
\begin{equation}\label{43}
{\bf J}_{+-} (\eta, \xi ) := {\rm i} \left(\begin{matrix}
- \eta \eta^+ & \eta \xi^+ \\
-\xi \eta^+ & \xi \xi^+
\end{matrix}\right) ,
\end{equation}
respectively,
where $\eta, \xi \in \mathbb{C}^n$ and $f,g \in C^\infty \big( \mathbb{C}^n \oplus \mathbb{C}^n\big)$. One has the following identity
\begin{equation*}
{\bf J}_{+-} (\eta , \xi )^2 = \big(\eta^+\eta - \xi^+\xi\big)\cdot {\bf J}_{+-} (\eta , \xi)
\end{equation*}
for this momentum map.
Hence, the momentum map ${\bf J}_{+-}$ maps the space of null-twistors $\mathcal{T}_{+-}^0 := I_{+-}^{-1} (0)$, where
\begin{equation*}
I_{+-} := \eta^+\eta - \xi^+ \xi ,
\end{equation*}
onto the nilpotent coadjoint orbit $\mathcal{N}_{1,0} = {\bf J}_0 (\mathcal{O}_{1,0})$ corresponding to $k=1$ and $l=0$ in sense of the classification presented in Proposition~\ref{prop:3}. The Hamiltonian flow $\sigma_{+-}^t \colon \mathbb{C}^n \oplus \mathbb{C}^n \to \mathbb{C}^n \oplus \mathbb{C}^n $, $t\in \mathbb{R}$, defined by $I_{+-}$ is given by
\begin{equation}\label{45}
\sigma_{+-}^t \left(\begin{matrix}
\eta \\
\xi
\end{matrix}\right) := {\rm e}^{{\rm i}t} \left(\begin{matrix}
\eta \\
\xi
\end{matrix}\right) .
\end{equation}

In order to describe fibre bundle structures of $\mathcal{N}_{1,0}\cong\mathcal{T}_{+-}^0 / U(1)$ we define the diffeomorphism $\Phi \colon \mathcal{T}_{+-}^0 \stackrel{\sim}{\rightarrow} \mathbb{S}^{2n-1}\times \dot{\mathbb{C}}^n$ by
\begin{equation*}
\Phi (\eta, \xi ) := \big((\eta^+\eta)^{-\frac{1}{2}} \eta , (\xi^+\xi)^{\frac{1}{2}} \xi \big) = (\eta' , \xi '),
\end{equation*}
where $\dot{\mathbb{C}}^n := \mathbb{C}^n \backslash \{0\}$. Note that $U(1)$ acts on $\mathcal{T}_{+-}^0$ as in~(\ref{45}).
The inverse diffeomorphism $\Phi^{-1}\colon \mathbb{S}^{2n-1}\times \dot{\mathbb{C}}^n \to \mathcal{T}_{+-}^0$ is given by
\begin{equation*}
\Phi^{-1} (\eta ' , \xi ' ) = \big((\xi '^+ \xi ' )^{\frac{1}{4}} \eta ' , (\xi '^+ \xi ' )^{-\frac{1}{4}} \xi ' \big).
\end{equation*}
These diffeomorphisms commute with the actions of Hamiltonian flow~(\ref{45}) on $\mathcal{T}_{+-}^0$ and on $\mathbb{S}^{2n-1}\times \dot{\mathbb{C}}^n$ which are defined by $(\eta, \xi )\mapsto (\lambda \eta , \lambda \xi )$ and by $(\eta ' , \xi ' ) \mapsto (\lambda \eta ' , \lambda \xi ')$, respectively, where $\lambda ={\rm e}^{{\rm i}t}$, $t\in \mathbb{R}$.

\begin{Proposition}\label{prop:41}\quad
\begin{enumerate}\itemsep=0pt
\item[$(i)$] Nilpotent orbit $\mathcal{N}_{1,0}$ is the total space of the fibre bundle
\begin{equation*}\unitlength=5mm\begin{picture}(11,4.6)
 \put(1,4){\makebox(0,0){$\mathbb{S}^{2n-1}$}}
 \put(8,4){\makebox(0,0){$\mathcal{N}_{1,0}$}}
 \put(9,0){\makebox(0,0){$\dot{\mathbb{C}}^n/ U(1) $}}
 \put(8.2,3){\vector(0,-1){2}}
 \put(2.5,4){\vector(1,0){3.7}}
\end{picture}\end{equation*}
over $\dot{\mathbb{C}}^n / U(1) $ with $\mathbb{S}^{2n-1}$ as a typical fibre. So, this bundle is a bundle of $(2n-1)$-dimensional spheres associated to $U(1)$-principal bundle $\dot{\mathbb{C}}^n \to \dot{\mathbb{C}}^n/ U(1) $.
\item[$(ii)$] One can also consider $\mathcal{N}_{1,0}$ as the total space of the fibre bundle
\begin{equation*}\unitlength=5mm\begin{picture}(11,4.6)
 \put(1,4){\makebox(0,0){$\dot{\mathbb{C}}^n$}}
 \put(8,4){\makebox(0,0){$\mathcal{N}_{1,0}$}}
 \put(9,0){\makebox(0,0){$\mathbb{C}\mathbb{P}(n-1) $}}
 \put(8.2,3){\vector(0,-1){2}}
 \put(2.5,4){\vector(1,0){3.7}}
 \end{picture}\end{equation*}
over complex projective space $\mathbb{C}\mathbb{P}(n-1)$ which is the base of Hopf $U(1)$-principal bundle $\mathbb{S}^{2n-1}\to \mathbb{S}^{2n-1}/U(1) \cong \mathbb{C}\mathbb{P}(n-1)$.
\end{enumerate}
\end{Proposition}

The total space of the tangent bundle $T\mathbb{C}\mathbb{P}(n-1) \to \mathbb{C}\mathbb{P}(n-1)$ has the form
\begin{equation*}
T\mathbb{C}\mathbb{P} (n-1) \cong \big\{(\eta ' , \xi ' ) \in \mathbb{S}^{2n-1} \times \mathbb{C}^n \colon \eta'^+ \xi ' =0 \big\} / U(1).
\end{equation*}
So, $T\mathbb{C}\mathbb{P}(n-1) \to \mathbb{C}\mathbb{P}(n-1)$ is vector subbundle of the vector bundle $\frac{\mathbb{S}^{2n-1} \times \mathbb{C}^n }{U(1)} \to \mathbb{S}^{2n-1}/U(1) \cong \mathbb{C}\mathbb{P}(n-1)$ and its complementary subbundle
\begin{equation*}
\mathbb{E} := \big\{ (\eta ' , \xi ' ) \in \mathbb{S}^{2n-1}\times \mathbb{C}^n \colon \xi' = s\eta' , s\in \mathbb{C}\big\}\to \mathbb{C}\mathbb{P}(n-1)
\end{equation*}
is isomorphic to the trivial bundle $\mathbb{C}\mathbb{P}(n-1) \times \mathbb{C}$.

Summing the above facts we conclude from the point (ii) of Proposition~\ref{prop:41} that one can identify $\mathcal{N}_{1,0}\cong \frac{\mathbb{S}^{2n-1} \times \mathbb{C}^n}{U(1)}\to \mathbb{C}\mathbb{P}(n-1)$ with the vector bundle $\frac{\mathbb{S}^{2n-1} \times \mathbb{C}^n}{U(1)}\to \mathbb{C}\mathbb{P}(n-1)$ with null section removed.

To explain the role of $U(n,n)$ as the dynamical group for $(2n-1)$-dimensional regularized Kepler problem we discuss now other description of $\mathcal{N}_{1,0}$ corresponding to the choice of anti-diagonal
\begin{equation*}
\phi_a := {\rm i}\left(\begin{matrix}
0 & -E \\
E & 0
\end{matrix}\right) ,
\end{equation*}
realization of twistor form (\ref{sproduct}). For diagonal realization $\phi_d$ see~(\ref{eq:312n}). Subsequently we will denote the realizations $\big(\mathbb{C}^{2n}, \phi_d \big)$ and $\big(\mathbb{C}^{2n}, \phi_a \big)$ of twistor space by $\mathcal{T}$ and $\tilde{\mathcal{T}}$, respectively. The same convention will be assumed for their groups of symmetry, i.e., $g=\left(\begin{smallmatrix} A & B \\ C& D \end{smallmatrix}\right) \in U(n,n)$ iff $g^+ \phi_d g = \phi_d$ and $\tilde{g} = \left(\begin{smallmatrix} \tilde{A}& \tilde{B}\\ \tilde{C} & \tilde{D} \end{smallmatrix}\right) \in \widetilde{U(n,n)}$ iff $\tilde{g}^+ \phi_a \tilde{g} = \phi_a $. Hence, for $\tilde{g} \in \widetilde{U(n,n)} $ and $\tilde{\mathfrak{X}}=\widetilde{{\mathfrak u}(n,n)}$ one has
\begin{gather*}
\tilde{A}^+\tilde{C} = \tilde{C}^+ \tilde{A}, \qquad
\tilde{D}^+ \tilde{B} = \tilde{B}^+ \tilde{D} ,\qquad
\tilde{A}^+ \tilde{D} = E+ \tilde{C}^+ \tilde{B}
\end{gather*}
and
\begin{equation*}
\tilde{\mathfrak{X}}
= \left(\begin{matrix} \tilde{\alpha}& \tilde{\beta}\\ \tilde{\gamma} & -\tilde{\alpha}^+ \end{matrix}\right),\end{equation*}
respectively, where $\tilde{\beta}^+=\tilde{\beta}$ and $\tilde{\gamma}^+=\tilde{\gamma}$.

The canonical one-form (\ref{41}) and the momentum map (\ref{43}) for $\tilde{\mathcal{T}}$ are given by
\begin{equation*}
\tilde{\gamma}_{+-} = \upsilon ^+ {\rm d}\zeta - \zeta^+ {\rm d}\upsilon
\end{equation*}
and by
\begin{equation*}
\tilde{{\bf J}}_{+-} (\upsilon , \zeta) = \left(\begin{matrix}
\upsilon \zeta^+ & - \upsilon \upsilon^+\\
\zeta\zeta^+ & - \zeta \upsilon^+ \end{matrix}\right),
\end{equation*}
where $\left(\begin{smallmatrix} \upsilon \\ \zeta \end{smallmatrix}\right) \in \tilde{\mathcal{T}}$. The null twistors space is defined as $\tilde{\mathcal{T}}_{+-}^0 := \tilde{I}_{+-}^{-1} (0) $, where
\begin{equation*}
\tilde{I}_{+-} (\upsilon , \zeta ) := {\rm i}\big(\zeta^+ \upsilon - \upsilon^+ \zeta \big).
\end{equation*}
The Hamiltonian flow on $\mathbb{C}^{2n} $ generated by $\tilde{I}_{+-}$ is given by
\begin{equation}\label{418}
\tilde{\sigma}_{+-}^t \left(\begin{matrix}
\upsilon \\
\zeta
\end{matrix}\right) = {\rm e}^{{\rm i}t} \left(\begin{matrix}
\upsilon \\
\zeta
\end{matrix}\right) \in \tilde{\mathcal{T}} .
\end{equation}

Both realizations $\mathcal{T}$ and $\tilde{\mathcal{T}}$ of the twistor space are related by the following unitary transformation of $\mathbb{C}^{2n}$:
\begin{equation*}
\left(\begin{matrix}
\upsilon \\
\zeta
\end{matrix}\right) = \mathcal{C}^+ \left(\begin{matrix}
\eta \\
\xi
\end{matrix}\right) \qquad \text{and} \qquad \left(\begin{matrix}
\eta \\
\xi
\end{matrix}\right) =\mathcal{C} \left(\begin{matrix}
\upsilon \\
\zeta
\end{matrix}\right) ,
\end{equation*}
where
\begin{equation}\label{420}
\mathcal{C}:= \frac{1}{\sqrt{2}}\left(\begin{matrix}
E & -{\rm i}E \\
-{\rm i}E & E \end{matrix}\right),
\end{equation}
which gives an isomorphism between the $U(n,n)$-Hamiltonian spaces $(\mathcal{T}, {\rm d}\gamma_{+-}) $ and $\big(\tilde{\mathcal{T}}, {\rm d}\tilde\gamma_{+-}\big) $.

Now let us consider $H(n)\times H(n)$ with $d\tilde{\gamma}_0$, where
\begin{equation*}
\tilde{\gamma}_0 :=\operatorname{Tr} (Y{\rm d}X)
\end{equation*}
and $(Y,X) \in H(n) \times H(n)$, as a symplectic manifold. We define the symplectic action of $\tilde{g} = \left(\begin{smallmatrix} \tilde{A}& \tilde{B}\\ \tilde{C} & \tilde{D} \end{smallmatrix}\right)$ on $ H(n) \times H(n) $ by
\begin{equation}\label{422}
\tilde{\sigma}_{\tilde{g}} (Y,X) := \big(\big(\tilde{A}Y+\tilde{B}\big)\big(\tilde{C}Y+\tilde{D}\big)^{-1},\big(\tilde{C}Y+\tilde{D}\big)X\big(\tilde{C}Y+\tilde{D}\big)^+ \big).
\end{equation}
Let us note here that the above action is not defined globally, i.e., the formula~(\ref{422}) is valid only if $\det \big(\tilde{C}Y + \tilde{D}\big) \neq 0$.

The momentum map $\tilde{{\bf J}}_0\colon H(n)\times H(n) \to \widetilde{{\mathfrak u}(n,n)} $ corresponding to ${\rm d}\tilde{\gamma}_0$ and $\tilde{\sigma}_{\tilde{g}}$ has the form
\begin{gather*}
\tilde{{\bf J}}_{0} (Y, X) = \left(\begin{matrix} YX & -YXY \\
X & -XY
\end{matrix}\right)
\end{gather*}
and it satisfies the equivariance property
\begin{equation*}
\tilde{{\bf J}} \circ \tilde{\sigma}_{\tilde{g}} = \operatorname{Ad}_{\tilde{g}} \circ \tilde{{\bf J}}.
\end{equation*}
The following diagram
\begin{equation} \unitlength=5mm\label{diagram}
\hspace*{6mm}\begin{split} &\begin{picture}(11,4.6)
 \put(1,4){\makebox(0,0){$T^*U(n)$}}
 \put(8,4){\makebox(0,0){${\mathfrak u}(n,n)$}}
 \put(15,4){\makebox(0,0){$\mathcal{T}$}}
 \put(15,0){\makebox(0,0){$\tilde{\mathcal{T}}$}}
 \put(8,0){\makebox(0,0){$\widetilde{{\mathfrak u}(n,n)}$}}
 \put(1,0){\makebox(0,0){$H(n)\times H(n)$}}
 \put(1,1){\vector(0,1){2}}
 \put(8,1){\vector(0,1){2}}
 \put(15,1){\vector(0,1){2}}
 \put(14,4){\vector(-1,0){4}}
 \put(14,0){\vector(-1,0){4}}
 \put(3.8,0){\vector(1,0){2.5}}
 \put(3,4){\vector(1,0){3}}
 \put(4.5,4.4){\makebox(0,0){${\bf J}_0$}}
 \put(11.5,4.4){\makebox(0,0){${\bf J}_{+-}$}}
 \put(11.5,0.5){\makebox(0,0){$\tilde{{\bf J}}_{+-}$}}
\put(4.5,0.5){\makebox(0,0){$ \tilde{{\bf J}}_0$}}
 \put(2,2){\makebox(0,0){$T^*_{\mathcal{C}}$}}
 \put(9,2){\makebox(0,0){$\operatorname{Ad}_\mathcal{C}$}}
\put(16,2){\makebox(0,0){$ \mathcal{C}$}}
\put(1.21,1){\makebox(0,0){$\cup$}}
 \end{picture} \end{split}\end{equation}
depicts relationship between the Poisson manifolds defined above. The maps represented by vertical arrows in (\ref{diagram}) are defined by (\ref{420}) and by
\begin{gather}
\operatorname{Ad}_\mathcal{C} \big(\tilde{\mathfrak{X}}\big) := \mathcal{C} \tilde{\mathfrak{X}} \mathcal{C}^+ ,\nonumber\\
\label{426}
T^*_\mathcal{C} (Y, X) := \left((Y-{\rm i}E)(-{\rm i}Y+E)^{-1}, \frac{{\rm i}}{2} (-{\rm i}Y+E)X(-{\rm i}Y+E)^+\right),
\end{gather}
where $\tilde{\mathfrak{X}} \in \widetilde{{\mathfrak u}(n,n)}$ and $(Y,X)\in H(n)\times H(n)$.

\begin{Proposition}\label{prop:42}
All arrows in the diagram \eqref{diagram} are the $U(n,n)$-equivariant Poisson maps.
\end{Proposition}
\begin{proof}
By straightforward verification.
\end{proof}

The first component in (\ref{426}), i.e.,
\begin{equation*}
Z= (Y-{\rm i}E)(-{\rm i}Y+E)^{-1}
\end{equation*}
is a smooth one-to-one map of $H(n)$ into $U(n)$, which is known as Cayley transform, see, e.g.,~\cite{faraut}. Hence, the unitary group~$U(n)$ could be considered as a compactification of~$H(n)$, Namely, in order to obtain the full group $U(n)$ one adds to Cayles image of~$H(n)$ such unitary matrices~$Z$, which satisfy the condition $\det({\rm i}Z+E)=0$. One sees this by observing that the inverse Cayley map is defined by
\begin{equation*}
Y=(Z+{\rm i}E)({\rm i}Z+E)^{-1},
\end{equation*}
if $\det({\rm i}Z+E) \neq 0$.

Let us define
\begin{equation*}
\dot{\tilde{\mathcal{O}}}_{1,0} := H(n) \times C_{n,1},
\end{equation*}
where
\[C_{n,1}:= \{ X\in H(n)\colon \dim (\operatorname{Im} (X)) =1 \ \text{and} \ X\geq 0 \},\]
and note that one has
\begin{gather}\label{435}
\tilde{{\bf J}}_0 \big(\dot{\tilde{\mathcal{O}}}_{1,0}\big) \subset \tilde{\mathcal{N}}_{1,0} = \tilde{{\bf J}}_{+-}\big(\tilde{\mathcal{T}}_{+-}^0 \big),\\
\label{433}
{\bf J}_0 (\mathcal{O}_{1,0}) = \mathcal{N}_{1,0} = {\bf J}_{+-}\big(\mathcal{T}_{+-}^0\big).
\end{gather}
Let us recall that positivity $X\geq 0$ of $X\in H(n)$ means the positivity of its eigenvalues. We mention here that in~\cite{meng5} the cotangent bundle~$T^*C_{n,1}$ of $C_{n,1}\subset H(n)$ is used as the phase space of generalized $U(1)$-Kepler problem.

Taking into account the properties of Poisson maps presented in the diagram (\ref{diagram}), as well as the relations (\ref{433}) and (\ref{435}), one obtains the following morphisms of the reduced $U(n,n)$-Hamiltonian spaces
\begin{equation}\unitlength=5mm \label{equivalence}
\hspace*{1mm}\begin{split} \begin{picture}(11,4.6)
 \put(1,4){\makebox(0,0){$\mathcal{O}_{1,0}/_\sim$}}
 \put(8,4){\makebox(0,0){$\mathcal{N}_{1,0}$}}
 \put(15.5,4){\makebox(0,0){$\mathcal{T}_{+-}^0/ _\sim$}}
 \put(15.5,0){\makebox(0,0){$\tilde{\mathcal{T}}_{+-}^0/ _\sim ,$}}
 \put(8,0){\makebox(0,0){$\tilde{\mathcal{N}}_{1,0}$}}
 \put(1,0){\makebox(0,0){$\dot{\tilde{\mathcal{O}}}_{1,0}/_\sim $}}
 \put(1,1){\vector(0,1){2}}
 \put(8,1){\vector(0,1){2}}
 \put(15,1){\vector(0,1){2}}
 \put(14,4){\vector(-1,0){4}}
 \put(14,0){\vector(-1,0){4}}
 \put(3.2,0){\vector(1,0){3}}
 \put(3,4){\vector(1,0){3}}
 \put(4.5,4.6){\makebox(0,0){${\bf J}_0/_\sim$}}
 \put(11.5,4.6){\makebox(0,0){${\bf J}_{+-}/_\sim$}}
\put(5,0.6){\makebox(0,0){$\tilde{{\bf J}}_0/_\sim$}}
 \put(11.5,0.6){\makebox(0,0){$\tilde{{\bf J}}_{+-}/_\sim$}}
\put(3.2,0.21){\makebox(0,0){$ \subset$}}
 \put(2.5,2){\makebox(0,0){$T^*_{\mathcal{C}}/_\sim$}}
 \put(9.5,2){\makebox(0,0){$\operatorname{Ad}_\mathcal{C} /_\sim$}}
\put(16.5,2){\makebox(0,0){$ \mathcal{C}/_\sim$}}
\put(0.79,1){\makebox(0,0){$\cup$}}
 \end{picture}\end{split}\end{equation}
which are symplectic isomorphisms, except for
\[
T^*_{\mathcal{C}}/_\sim\colon \ \dot{\tilde{\mathcal{O}}}_{1,0}/_\sim \hookrightarrow \mathcal{O}_{1,0}/_\sim \qquad \text{and} \qquad \tilde{{\bf J}}_0/_\sim\colon \ \dot{\tilde{\mathcal{O}}}_{1,0}/_\sim \hookrightarrow \tilde{\mathcal{N}}_{1,0},
\]
 which are one-to-one symplectic maps only. The equivalence relations $\sim$ in (\ref{equivalence}) are defined by the degeneracy leaves of the restrictions of respective symplectic forms defined on manifolds which appear on the left- and right-hand sides of the diagram~(\ref{diagram}).

For any $\mathfrak{X}=\left(\begin{smallmatrix} a & b \\ b^+ & d \end{smallmatrix}\right) \in {\mathfrak u}(n,n)$ one defines the linear function
\begin{equation}\label{438}
L_\mathfrak{X} \left(\begin{matrix}
\alpha & \beta \\
\beta^+ & \delta \end{matrix}\right) := \operatorname{Tr}\left(\left(\begin{matrix} a & b \\ b^+ & d\end{matrix}\right)\left(\begin{matrix} \alpha & \beta \\ \beta^+ & \delta \end{matrix}\right)\right)
\end{equation}
on the Lie--Poisson space $({\mathfrak u}(n,n), \{\cdot, \cdot \}_{\text{\rm L-P}})	$, where the Lie--Poisson bracket $\{\cdot, \cdot \}_{\text{\rm L-P}}$ is defined in~(\ref{eq:325}). These functions satisfy
\begin{equation*}
\left\{ L_{\mathfrak{X}_1} , L_{\mathfrak{X}_2}\right\}_{\text{L-P}} = L_{[\mathfrak{X}_1,\mathfrak{X}_2]}.
\end{equation*}
In particular cases when $\mathfrak{X}_{++}= {\rm i}\left(\begin{smallmatrix} E & 0 \\ 0 & E \end{smallmatrix}\right)$ and $\mathfrak{X}_{+-}= {\rm i}\left(\begin{smallmatrix} E & 0 \\ 0 & -E \end{smallmatrix}\right)$ one obtains
\begin{gather}
\label{440}
\big(L_{\mathfrak{X}_{++}} \circ {\bf J}_{+-}\big) (\eta , \xi ) = \eta^+\eta - \xi^+\xi = I_{+-},\\
\label{438n}
\big(L_{\mathfrak{X}_{+-}} \circ {\bf J}_{+-}\big) (\eta , \xi ) = \eta^+\eta + \xi^+\xi =:I_{++},\\
\label{439n}
\big(L_{\mathfrak{X}_{+-}} \circ {\bf J}_{0}\big) (Z , \rho ) = -2{\rm i}\operatorname{Tr} \rho =: I_0, \\
\big(L_{\mathfrak{X}_{++}} \circ {\bf J}_{0}\big) (Z , \rho ) = 0.
\end{gather}
Rewriting the above formula in the anti-diagonal realization, where $\tilde{\mathfrak{X}}_{++}=\mathcal{C} \mathfrak{X}_{++} \mathcal{C}^+= \mathfrak{X}_{++}$ and $\tilde{\mathfrak{X}}_{+-} =\left(\begin{smallmatrix} 0 & -E \\ E & 0 \end{smallmatrix}\right)= \mathcal{C} {\mathfrak{X}}_{+-}\mathcal{C}^+$ we find
\begin{gather}
\big(L_{\tilde{\mathfrak{X}}_{++}} \circ \tilde{{\bf J}}_{+-}\big) (\upsilon , \zeta ) = {\rm i}\big(\upsilon \zeta^+ - \zeta \upsilon^+\big),\\
\label{442}
\big(L_{\tilde{\mathfrak{X}}_{+-}} \circ \tilde{{\bf J}}_{+-}\big) (\upsilon , \zeta ) = \upsilon^+\upsilon + \zeta^+\zeta =:\tilde{I}_{++},\\
\label{447}
\big(L_{\tilde{\mathfrak{X}}_{+-}} \circ \tilde{{\bf J}}_{0}\big) (Y, X ) = \operatorname{Tr} \big(X\big(E+Y^2\big)\big) =: \tilde{I}_0,\\
\big(L_{\tilde{\mathfrak{X}}_{++}} \circ \tilde{{\bf J}}_{0}\big) (Y , X ) = 0.
\end{gather}
The functions $I_{++}$, $I_0$, and $\tilde{I}_{++}$, $\tilde{I}_0$ are invariants of the Hamiltonian flows presented in~(\ref{45})
and~(\ref{418}), respectively. Note that these flows are generated by $\mathfrak{X}_{++}= \tilde{\mathfrak{X}}_{++}\in {\mathfrak u}(n,n)\cap \widetilde{{\mathfrak u}(n,n)}$. So, the reduced functions $I_{++}/_\sim$, $I_{0}/_\sim$, $\tilde{I}_{++}/_\sim$ and $\tilde{I}_0/_\sim$ defined by (\ref{438n}), (\ref{439n}), (\ref{442}) and~(\ref{447}), respectively, could be considered as Hamiltonians on the reduced symplectic manifolds $\mathcal{T}_{+-}^0/_\sim$, $\mathcal{O}_{1,0}/_\sim $, $\tilde{\mathcal{T}}_{+-}^0/_\sim$ and $\dot{\tilde{\mathcal{O}}}_{1,0}/_\sim $. The function $L_{\mathfrak{X}_{+-}}\colon {\mathfrak u}(n,n) \to \mathbb{R}$, see definition~(\ref{438}), as well as the function $L_{\tilde{\mathfrak{X}}_{+-}} \colon \widetilde{{\mathfrak u}(n,n)}\to \mathbb{R}$, after restriction to $\mathcal{N}_{1,0}\subset {\mathfrak u}(n,n)$ and to $\tilde{\mathcal{N}}_{1,0} \subset \widetilde{{\mathfrak u}(n,n)}$ give Hamiltonians on $\mathcal{N}_{1,0}$ and on $\tilde{\mathcal{N}}_{1,0}$, respectively. Taking into account the symplectic manifolds morphisms mentioned in the diagram~(\ref{equivalence}) we conclude
\begin{Proposition}\label{prop:nowe}\quad
\begin{enumerate}\itemsep=0pt
\item[$(i)$] The $U(n,n)$-Hamiltonian systems: $\big(\mathcal{T}_{+-}^0/_\sim , I_{++}/_\sim\big)$, $\big(\tilde{\mathcal{T}}_{+-}^0/_\sim , \tilde{I}_{++}/_\sim\big)$, $(\mathcal{O}_{1,0}/_\sim , I_0/_\sim)$, \linebreak $(\mathcal{N}_{1,0}, L_{{\mathfrak{X}}_{+-}})$ and $\big(\tilde{\mathcal{N}}_{1,0}, L_{\tilde{\mathfrak{X}}_{+-}}\big)$ are mutually isomorphic.
\item[$(ii)$] The Hamiltonian system $\big(\dot{\tilde{\mathcal{O}}}_{1,0}/_\sim , \tilde{I}_0/_\sim\big)$ possesses two extensions $($regularizations$)$ to \linebreak $U(n,n)$-Hamiltonian systems given by the injective symplectomorphisms $T^*_\mathcal{C}/_\sim\colon \dot{\tilde{\mathcal{O}}}_{1,0}/_\sim \hookrightarrow \mathcal{O}_{1,0}/_\sim $ and $\tilde{{\bf J}}_0/_\sim\colon \dot{\tilde{\mathcal{O}}}_{1,0}/_{\sim}\hookrightarrow \tilde{\mathcal{N}}_{1,0}$, respectively.
\end{enumerate}
\end{Proposition}

Since the Hamiltonian $L_{\mathfrak{X}_{+-}}$ and, thus Hamiltonians $I_{++}/_\sim$, $I_0/_\sim$, $\tilde{I}_{++}/_\sim$ and $\tilde I_0/_\sim$ are defined by the element $\mathfrak{X}_{+-}$ of the Lie algebra ${\mathfrak u}(n,n)$ one can consider $U(n,n)$ as a dynamical group for all systems mentioned in (i) of Proposition~\ref{prop:nowe}. As a matter of fact we can treat all of them as various realizations of the same Hamiltonian system. See also \cite{iwai2,iwai3,toshihiro} for $U(n,n)$ as the dynamical group of MIC-Kepler system.

The easiest way to find the symmetry groups of these systems, and thus, their integrals of motion, is to consider the case $\big(\mathcal{T}_{+-}^0/_\sim , I_{++}/_\sim\big)$. In this case the symmetry group is the subgroup of $U(n,n)$, which preserve the canonical form $\gamma_{+-}$, defined in~(\ref{41}), and the Hamiltonian~$I_{++}$, i.e., it is $U(n,n)\cap U(2n) \cong U(n)\times U(n)$. So, the corresponding integrals of motion one obtains by restricting the matrix functions
\begin{equation*}
I_+ \big( \eta^+, \xi^+,\eta , \xi\big):= \eta\eta^+ \qquad \text{and} \qquad I_-\big( \eta^+, \xi^+, \eta, \xi \big):= \xi\xi^+
\end{equation*}
to $\mathcal{T}_{+-}^0$. Let us note that $\{I_{++}, I_+\}_{+-} = \{I_{++}, I_-\}_{+-} = 0$.

The integrals of motion $M\colon H(n)\times H(n) \to H(n)$ and $R\colon H(n)\times H(n)\to H(n)$ for Hamiltonian system $(H(n)\times H(n), \tilde{I}_0)$ have the following matrix forms
\begin{equation}\label{449}
M:= {\rm i}[X, Y]\qquad \text{and}\qquad R:= X + YXY.
\end{equation}

Reducing them to $\big(\dot{\tilde{\mathcal{O}}}_{1,0}/_\sim , \tilde{I}_0/_\sim\big)$ we obtain their correspondence
\begin{equation}\label{516}
I_+ \circ (C/_\sim)\circ \mathcal{K}_{\rm reg} = \frac{1}{2} (R-M) \qquad\text{and}\qquad I_- \circ(C/_\sim)\circ \mathcal{K}_{\rm reg} = \frac{1}{2} (R+M)
\end{equation}
to the integrals of motion $I_+$ and $I_-$, where $\mathcal{K}_{\rm reg}\colon \dot{\tilde{\mathcal{O}}}_{1,0}/_\sim\to \tilde{\mathcal{T}}_{+-}^0/_\sim$ is defined by
\begin{equation}\label{K}
\mathcal{K}_{\rm reg}:=\big(\tilde{{\bf J}}_{+-}/_\sim\big)^{-1}\circ \big(\tilde{{\bf J}}_{0}/_\sim\big)
\end{equation}

The Hamilton equations defined by $\tilde{I}_0$ are
\begin{equation}\label{450}
\frac{{\rm d}}{{\rm d}t} Y = E + Y^2,\qquad \frac{{\rm d}}{{\rm d}t} X = -(XY-YX),
\end{equation}
i.e., they could be classified as a matrix Riccati type equations. In order to obtain their solution we note that after passing to $\big(\mathcal{T}_{+-}^0/_\sim , I_{++}/_\sim\big)$ they asssume the form of a linear equations which are solved by
\begin{equation*}
\sigma_{ ++}^t \left(\begin{matrix}
\eta\\
\xi \end{matrix}\right) = \left(\begin{matrix}
{\rm e}^{{\rm i}t}E & 0 \\
0 & {\rm e}^{-{\rm i}t}E
\end{matrix}\right)\left(\begin{matrix}
\eta\\
\xi
\end{matrix}\right),
\end{equation*}
i.e., the Hamiltonian flow $\sigma_{++}^t$ is one-parameter subgroup of $U(n,n)$ generated by $ \mathfrak{X}_{+-} \in {\mathfrak u}(n,n)$. Therefore, going through the symplectic manifold isomorphisms presented in~(\ref{equivalence}), we obtain the solution
\begin{gather*}
\begin{split}
& Y(t) = (Y \cosh t - {\rm i} E\sinh t )({\rm i}Y\sinh t + E \cosh t)^{-1},\\
& X(t) = ({\rm i}Y\sinh t + E \cosh t)X ({\rm i}Y\sinh t + E \cosh t )^+
\end{split}
\end{gather*}
of (\ref{450}) by specifying the transformation formula (\ref{422}) to the one-parameter subgroup $\tilde\sigma_{+-}^t= \mathcal{C}^+ \left(\begin{smallmatrix}
{\rm e}^{{\rm i}t}E & 0 \\
0 & {\rm e}^{-{\rm i}t}E
\end{smallmatrix}\right) \mathcal{C}$ of the group $\widetilde{U(n,n)}$.

\looseness=-1 Ending this section let us mention the papers \cite{oh,mladenov2,mladenov,os,simms}, where Kepler and MIC-Kepler problems were considered on the classical and quantum levels. Let us also mention some interesting generalizations of these problems \cite{bm,meng,meng4,meng2,meng3} based on the theory of Jordan algebras.

\section{Cayley and Kustaanheimo--Stiefel transformations}\label{sec:Cayley}

In this section we discuss two regularizations of the Hamiltonian system $\big(\dot{\tilde{\mathcal{O}}}_{1,0}/_\sim,\tilde {I}_0\big)$ which were mentioned in the point (ii) of Proposition~\ref{prop:nowe}. At first we will show that the regularization $\mathcal{K}_{\rm reg}\colon\dot{\tilde{\mathcal{O}}}_{1,0}/_\sim\to \tilde{\mathcal{T}}_{+-}^0/_\sim $, defined in (\ref{K}), could be interpreted as a generalization for arbitrary dimension of Kustaanheimo--Stiefel regularization, which was introduced in \cite{stiefel} for the case $n=2$. Then we will discuss shortly the regularization $\mathcal{C}_{\rm reg}\colon \dot{\tilde{\mathcal{O}}}/_\sim\to \mathcal{T}^0_{+-}/_\sim$ defined by Cayley transformation
\begin{equation*}
\mathcal{C}_{\rm reg}:=({\bf J}_{+-}/_\sim)^{-1}\circ ({\bf J}_{0}/_\sim)\circ ({T}^*_C/_\sim).
\end{equation*}
We will also show the equivalence of the both considered regularizations.

Comparing the values
\begin{equation*}
\tilde{{\bf J}}_{0}(X,Y)=\left(\begin{matrix} YX & -YXY \\ X & -XY \end{matrix}\right)=
\left(\begin{matrix} v\zeta^+ & -vv^+ \\ \zeta\zeta^+ & -\zeta v^+ \end{matrix}\right)=\tilde{{\bf J}}_{+-}(v,\zeta)
\end{equation*}
 of momentum maps $\tilde{{\bf J}}_0$ and $\tilde{{\bf J}}_{+-}$ we find
that $(Y,X)\in \tilde{{\bf J}}_{0}^{-1}\big(\tilde{{\bf J}}_{+-}(v,\zeta)\big) $ iff
\begin{gather}\label{wzor2a}
X=\zeta\zeta^+,
\\ \label{wzor2}
v=Y\zeta.
\end{gather}

Let us define
$\dot{\tilde{\mathcal{T}}}^0_{+-}:=\big\{\left(\begin{smallmatrix} v\\ \zeta \end{smallmatrix}\right)\in \tilde{\mathcal{T}}^0_{+-}\colon \zeta\not=0\big\}$
and observe that the surjective submersion $\mathcal{R}\colon \dot{\tilde{\mathcal{O}}}_{1,0}\to \dot{\tilde{\mathcal{T}}}_{+-}^0/_\sim$ defined by
\begin{equation*}
\mathcal{R}(Y,X):=\left[\left(\begin{matrix} Y\zeta \\ \zeta\end{matrix}\right)\right],
\end{equation*}
where $X=\zeta\zeta^+$ and $\left[\left(\begin{smallmatrix} Y\zeta \\ \zeta\end{smallmatrix}\right)\right]:=\left\{{\lambda}\left(\begin{smallmatrix} Y\zeta \\ \zeta\end{smallmatrix}\right)\colon \lambda\in U(1)\right\}$, satisfies
\begin{gather*}
\mathcal{R}_*\tilde{\gamma}_{+-}|_{\dot{\tilde{\mathcal{T}}}^0_{+-}}=\tilde{\gamma}_{0}|_{\dot{\tilde{\mathcal{O}}}_{1,0}},
\qquad
\tilde I_{++}\circ \mathcal{R}=\tilde I_{0}.
\end{gather*}

We also observe that the fibres $\mathcal{R}^{-1}\left(\left[\left(\begin{smallmatrix} Y\zeta \\ \zeta\end{smallmatrix}\right)\right]\right)$, where $\left[\left(\begin{smallmatrix} Y\zeta \\ \zeta\end{smallmatrix}\right)\right]\in \dot{\tilde{\mathcal{T}}}^0_{+-}/_\sim$ are the degeneracy leaves of ${\rm d}\tilde{\gamma}_0/_{\dot{\tilde{\mathcal{O}}}_{1,0}}$, so, one can identify the quotient map $\tilde{\mathcal{R}}\colon {\dot{\tilde{\mathcal{O}}}_{1,0}}/_\sim\to {\dot{\tilde{\mathcal{T}}}}_{+-}^0/_\sim$ with the map $\mathcal{K}_{\rm reg}\colon \dot{\tilde{\mathcal{O}}}_{1,0}/_\sim \to \dot{\tilde{\mathcal{T}}}_{+-}^0/_\sim$ defined in (\ref{K}).

In order to obtain explicitly a local expression for $\tilde{\mathcal{R}}^{-1}$ let us take the map $\mathcal{S}\colon \Omega\to \dot{\tilde{\mathcal{O}}}_{1,0}$ defined by
\begin{equation}\label{wzor6}
\mathcal{S}\big(v^+, \zeta^+, v,\zeta\big):=\big(Y\big(\upsilon^+,\zeta^+, \upsilon , \zeta\big), \zeta\zeta^+\big)\end{equation}
on an open $U(1)$-invariant subset $\Omega\subset \tilde{\mathcal{T}}_{+-}^0$, where the map $Y\colon \Omega\to H(n)$ fulfills the conditions
\begin{gather}\label{wzor7}
Y\big(\upsilon^+,\zeta^+, \upsilon , \zeta\big)\zeta=\upsilon,
\end{gather}
and
\begin{equation}\label{wzor8}
Y\big((\lambda \upsilon)^+,(\lambda \zeta)^+, \lambda \upsilon , \lambda \zeta\big)=Y\big(\upsilon^+,\zeta^+, \upsilon , \zeta\big)
\end{equation}
for $\lambda\in U(1)$. From (\ref{wzor7}) and (\ref{wzor8}) we see that $\mathcal{S}$ is a local section of $\mathcal{R}$, i.e., $\mathcal{R}\circ\mathcal{S}={\rm id}_\Omega$. Thus one can choose $\mathcal{S}\big(\upsilon^+,\zeta^+, \upsilon , \zeta\big)\in \mathcal {R}^{-1}(\left[\left(\begin{smallmatrix} \upsilon \\ \zeta\end{smallmatrix}\right)\right])$ as a representative of the degeneracy leaf
\[
\mathcal {R}^{-1} \left[\left(\begin{matrix} \upsilon \\ \zeta\end{matrix}\right)\right]=\big\{\big(Y(\upsilon^+,\zeta^+, \upsilon , \zeta)+Y',\zeta \zeta^+\big)\colon Y'\in H(n)\ {\rm and}\ Y'\zeta=0\big\}
\]
of the differential closed form ${\rm d}\tilde{\gamma}_0|_{\dot{\tilde{\mathcal{O}}}_{1,0}}$. Let ``$\sim$'' be the equivalence relation on $\dot{\tilde{\mathcal{O}}}_{1,0}$ defined by these leaves, then identifying the quotient manifold $\mathcal{R}^{-1}(\Omega/_\sim)/_\sim$, with respect to this equivalence, with the local section $\mathcal{S}(\Omega)$ we obtain the following local diffeomorphism ${\mathcal{S}}\colon \Omega/_\sim\stackrel{\sim}{\to} \mathcal{S} (\Omega)\cong \mathcal{R}^{-1}(\Omega/_\sim)/_\sim$.

In next examples we will present two local sections $S\colon \Omega \to \dot{\tilde{\mathcal{O}}}_{1,0}$ of $\mathcal{R}\colon \dot{\tilde{\mathcal{O}}}_{1,0}\to \dot{\tilde{\mathcal{T}}}_{+-}^0/_\sim$.

\begin{Example} Let us take $\Omega=\dot{\tilde{\mathcal{T}}}^0_{+-}$
and define $Y\colon \Omega\to H(n) $ as follows
\begin{equation}\label{wzor10}
Y\big(\upsilon^+,\zeta^+ , \upsilon , \zeta\big):=\frac{1}{\zeta^+\zeta}\left[\zeta \upsilon^++\upsilon\zeta^+-\frac{1}{2}\big(\upsilon^+\zeta+\zeta^+ \upsilon\big)E\right].\end{equation}

One easily checks that the map $Y\colon \Omega\to H(n)$ defined in~(\ref{wzor10}) satisfies the conditions (\ref{wzor7}) and (\ref{wzor8}), so, it defines by~(\ref{wzor6}) a local section of $\mathcal{R}$.
\end{Example}

\begin{Example} In this example we assume
$\Omega:=\big\{\left(\begin{smallmatrix} v \\ \zeta\end{smallmatrix}\right)\in \tilde{\mathcal{T}}^0_{+-}\colon v^+\zeta\not=0\big\}$ and define $Y\colon \Omega\to H(n)$ by
\begin{equation*}Y\big(\upsilon^+,\zeta^+ , \upsilon , \zeta\big)=\frac{\upsilon\upsilon^+}{\upsilon^+\zeta}.\end{equation*}
\end{Example}

The meaning of the first example will be explained at the end of this section. The second example illustrates another possibility to define a local diffeomorphism $\mathcal{S}\colon \Omega/_\sim\stackrel{\sim}{\to} \mathcal{S} (\Omega)\cong \mathcal{R}^{-1}(\Omega/_\sim)/_\sim$.

Having in mind a physical interpretations of the discussed Hamiltonian systems, we will consider the case $n=2$ in details.
 Expanding $(Y,X) \in H(2)\times H(2)$ in Pauli matrices $\sigma_0: = \left(\begin{smallmatrix} 1 & 0 \\ 0 & 1 \end{smallmatrix}\right)$, $\sigma_1 := \left(\begin{smallmatrix} 0 & 1 \\ 1 & 0 \end{smallmatrix}\right)$, $\sigma_2 := \left(\begin{smallmatrix} 0 & {\rm i} \\ -{\rm i} & 0 \end{smallmatrix}\right)$ and $\sigma_3 := \left(\begin{smallmatrix} 1 & 0 \\ 0 & -1 \end{smallmatrix}\right)$, i.e.,
\begin{equation}\label{456}
Y= y^0 \sigma_0+ \vec{y}\cdot\vec{ \sigma}\qquad\text{and} \qquad X = x^0 \sigma_0+ \vec{x}\cdot\vec{ \sigma},
\end{equation}
where $\vec{\sigma} = (\sigma_1, \sigma_2 , \sigma_3)$, we find that
\begin{equation*}
\frac 12 \tilde{\gamma}_{0} =y^0 {\rm d}x^0+ \vec{y}\cdot {\rm d}\vec{ x}.
\end{equation*}

In this case we assume that $\Omega=\dot{\tilde{\mathcal{T}}}_{+-}^0$ and define $\mathcal{S}\colon \dot{\tilde{\mathcal{T}}}_{+-}^0\to \dot{\tilde{\mathcal{O}}}_{1,0}$ taking $Y\colon \dot{\tilde{\mathcal{T}}}_{+-}^0\to H(n)$ such as in~\eqref{wzor10}. We see from~(\ref{wzor6}) and (\ref{wzor10}) that $(Y,X)\in \mathcal{S}\big(\dot{\tilde{\mathcal{T}}}_{+-}^0\big)$ iff $\operatorname{Tr}( Y)=2y^0=0$ and $\det X={x^0}^2-\vec{x}^2=0$, $\operatorname{Tr} (X)=2x^0>0$.
 From the above it follows that $\mathcal{S}\big(\dot{\tilde{\mathcal{T}}}_{+-}^0\big) \cong \mathbb{R}^3 \times \dot{\mathbb{R}}^3$, where $\dot{\mathbb{R}}^3={\mathbb{R}}^3\setminus\{0\}$, and the canonical form $\tilde{\gamma}_0$ after restriction to $\mathcal{S}\big(\dot{\tilde{\mathcal{T}}}_{+-}^0\big)$ is given by
\begin{equation*}
\tilde{\gamma}_{0}|_{\mathcal{S}(\dot{\tilde{\mathcal{T}}}_{+-}^0)} = 2\vec{y}\cdot {\rm d}\vec{x} = 2 y_k {\rm d}x^k.
\end{equation*}
Using the identity
\begin{equation}\label{460}
\sigma_k \sigma_l + \sigma_l \sigma_k = 2\delta_{kl}
\end{equation}
valid for Pauli matrices $\sigma_k$, $k=1,2,3$, we find that the Hamiltonian $H_0:= \frac{1}{2}\tilde{I}_0$, defined in~(\ref{447}), after restriction to $\mathcal{S}\big(\dot{\tilde{\mathcal{T}}}_{+-}^0\big)$ assumes the following form
\begin{equation*}
H_{0} =\tilde{I}_0|_{\mathcal{S}(\dot{\tilde{\mathcal{T}}}_{+-}^0)}= \|\vec{x} \| \big(1+ \|\vec{y} \|^2\big)
\end{equation*}
on $\mathbb R^3\times \dot{\mathbb R}^3$. Let us note that $ \|\vec{x} \| = x^0 = \frac{1}{2}\zeta^+ \zeta >0$.

Summing up the above facts we state that the Hamiltonian system $\big(H(2)\times H(2) , {\rm d}\tilde{\gamma}_{0}, \tilde{I}_0\big)$ after reduction to $\big(\mathbb R^3\times \dot{\mathbb R}^3, 2{\rm d}\vec{y} \wedge {\rm d}\vec{x}, H_0\big)$ is exactly the $3$-dimensional Kepler system written in the ``fictitious time''~$s$ which is related to the real time~$t$ via the rescaling
\begin{equation*}
\frac{{\rm d}s}{{\rm d}t} = \frac{1}{\|\vec{x}\|}.
\end{equation*}
For an exhaustive description of the regularized Kepler problem we address to original papers of Moser~\cite{moser} and of Kustaanheimo and Stiefel~\cite{stiefel} as well as to~\cite{kummer}, where the relationship between Moser and Kustaanheimo--Stiefel regularization was established.

In order to express $(\vec{y}, \vec{x}) \in \mathbb{R}^3 \times \dot{\mathbb{R}}^3$ by $\left(\begin{smallmatrix} \upsilon \\ \zeta \end{smallmatrix}\right) \in \tilde{\mathcal{T}}_{+-}^0 $ we put $Y= \vec{y}\cdot\vec{\sigma} = y_k \sigma_k$ into (\ref{wzor2}) and multiply this equation by $\zeta^+\sigma_l$. Then, using~(\ref{460}) and~(\ref{wzor2a}) we obtain the one-to-one map defined by
\begin{gather}\label{464}
\vec{y} = \frac{1}{\zeta^+\zeta } \frac{1}{2} \big(\upsilon^+ \vec{\sigma}\zeta + \zeta^+ \vec{\sigma} \upsilon\big), \qquad
\vec{x} = \frac{1}{2}\zeta^+ \vec{\sigma} \zeta ,
\end{gather}
of $\dot{\tilde{\mathcal{T}}}_{+-}^0 /_\sim $ onto $\mathbb{R}^3 \times \dot{\mathbb{R}}^3$.
This map is known in literature of celestial mechanics as Kustaanheimo--Stiefel transformation, see \cite{kummer,stiefel}. There are possible some variations of~(\ref{464}) naturally presented in quaternion language, see equation~(15) in~\cite{ferrer}. This quaternionic approach does not extend to an arbitrary dimension, where symplectic geometry methods are effective only.

Therefore, having in mind the case $n=2$, it is reasonable to interpret:
\begin{enumerate}\itemsep=0pt
\item[i)] the Hamiltonian systems $\big(\mathcal{T}^0_{+-}/_\sim,I_{++}\big)$, $\big(\tilde{\mathcal{T}}^0_{+-}/_\sim,\tilde{I}_{++}\big)$, $\big(\mathcal{O}_{1,0}/_\sim, I_0\big)$, $\big(\tilde{\mathcal{N}}_{1,0},L_{{\tilde{\mathfrak{X}}}_{+-}}\big)$ and $\big(\mathcal{N}_{1,0},L_{{\mathfrak{X}}_{+-}}\big)$ as the various equivalent realizations of the regularized $(2n-1)$-dimensional Kepler problem;
\item[ii)] the map $\mathcal{S}\colon \dot{\tilde{\mathcal{T}}}^0_{+-}\to \mathcal{S}\big(\dot{\tilde{\mathcal{T}}}^0_{+-}\big)$, where $Y\colon \dot{\tilde{\mathcal{T}}}^0_{+-}\to H(n)$ is given by~\eqref{wzor10}, as Kustaanheimo--Stiefel transformation for the $(2n-1)$-dimensional Kepler problem.
\end{enumerate}

Finally let us briefly discuss the regularization of $\big(\dot{\tilde{\mathcal{O}}}_{1,0}/_\sim,\tilde{I}_0\big)$ given by $\mathcal{C}_{\rm reg}\colon \dot{\tilde{\mathcal{O}}}_{1,0}/_\sim\to \mathcal{T}^0_{+-}/_\sim$ which we will call Cayley regularization of the $(2n-1)$-dimensional Kepler problem.
From the commutativity of the diagram \eqref{equivalence} we conclude that
\begin{equation*}
 \mathcal C_{\rm reg} = (\mathcal C/\sim)\circ \mathcal K_{\rm reg}.
\end{equation*}
Therefore, the Kustaanheimo--Stiefel regularization is equivalent to the Cayley regularization of the $(2n-1)$-dimensional Kepler problem.

By Proposition~\ref{prop:nowe} the $(2n-1)$-Kepler system $\big(\dot{\tilde{\mathcal{O}}}_{1,0} , \tilde{I}_0/_\sim\big)$ is extended to (regularized by) arbitrary $U(n,n)$-Hamiltonian system occurred in the diagram~(\ref{equivalence}). In accordance with terminology assumed here, the extension of $\big(\dot{\tilde{\mathcal{O}}}_{1,0} , \tilde{I}_0/_\sim\big)$ to a $U(n,n)$-Hamiltonian system from the upper row of the diagram (\ref{equivalence}) is called the Cayley regularization, whereas the extension to the one from the lower row is the Kustaanheimo--Stiefel regularization. The justification of this nomenclature follows from the appearance in~(\ref{equivalence}) the maps~(\ref{426}) and~(\ref{wzor6}).

The benefit of using the various isomorphic realizations of the same $U(n,n)$-Hamiltonian system is based on the possibility to admit different physical interpretations for them. For example, if $n=2$ one can consider the symplectic manifold $\mathcal{O}_{1,0}/_\sim $ as the phase space of massless scalar particle in the conformally compactified Minkowski space $\overline{M}_{1,3} \cong U(2)$, see \cite{AO}. The realizations $\tilde{\mathcal{T}}^0_{+-}/_\sim $ and $\mathcal{T}^0_{+-}/_\sim $ play the crucial role in the twistor theory \cite{penrose} of R.~Penrose.

In the papers \cite{cordani2, cordani} a method of linearization of the regularized Kepler problem based on the Clifford algebra $C(2, n+1)$ of the Lie group ${\rm SO}(2, n+1)$ was proposed. The ${\rm Spin}(2, n+2)$-invariant symplectic structure $\omega$ on an ideal $V\subset C(2, n+1)$ of the Clifford algebra $C(2, n+1)$ is fixed. Then, using the momentum map ${\bf J}\colon V\to \mathfrak{sl}(2, n+1)$ on this auxiliary ${\rm Sp}(2, n+2)$-symplectic manifold $(V, \omega)$, the Marsden--Weinstein reduction procedure to the $\operatorname{Ad}^* ({\rm Spin} (2, n+2))$-orbits $\mathcal{O}= \iota (T^* \mathbb{S}^n)$ is applied. The inverse $(KS)^{-1}$ of Kustaanheimo--Stiefel map is defined by $(KS)^{-1} = l \circ \pi$, where symplectomorphism $l$ is defined as the one making the diagram \begin{equation*}\unitlength=5mm 
\hspace*{5mm}\begin{picture}(11,4.6)
 \put(1,4){\makebox(0,0){$T^*\big(\mathbb{R}^n\backslash \{0\}\big)$}}
 \put(9,4){\makebox(0,0){$T^+\mathbb{S}^n$}}
 \put(15,4){\makebox(0,0){$\mathcal{O}$}}
 \put(8.5,0){\makebox(0,0){${\bf J}^{-1}(\mathcal{O})/\sim$}}
 \put(9,3){\vector(0,-1){2}}
 \put(4,4){\vector(1,0){3}}
		\put(11,4){\vector(1,0){3}}
		\put(11,0){\vector(1,1){3}}
\put(5.5,4.4){\makebox(0,0){$\pi$}}
\put(12.5,4.4){\makebox(0,0){$\iota$}}
 \put(10,2){\makebox(0,0){$l$}}
		\put(14,1.5){\makebox(0,0){${\bf J}/\sim$}}
\put(4,4.25){\makebox(0,0){$\subset$}}
 \end{picture}\end{equation*}
commutative, see \cite{cordani, efs}, where $\pi$ is Moser regularization \cite{moser} map and $\iota$ is the momentum map for Moser phase space $T^+\mathbb{S}^n = T^* \mathbb{S}^n \backslash \{\mbox{null section}\}$. Comparing the above approach with ours, we conclude that the construction of Kustaanheimo--Stiefel map presented in \cite{cordani, efs} combines the symplectic geometry with Clifford algebras theory and is obtained in an implicit way. In our case we use the Poisson geometry methods only and obtain the explicit formulas, see~(\ref{wzor6}),~(\ref{wzor10}), for Kustaanheimo--Stiefel map. Both approaches intersect in the case $n=2$.

Although here we have considered the odd-dimensional Kepler problem only, the even-dimensional case is none the less important. Since one can obtain the planar Kepler problem from the spatial one by some reduction procedure \cite{n1,n2}, the question arise: is it possible in general case? Another interesting question concerns the Kepler problem of positive energy. But, these are the tasks for a next paper.

\section[An integrable generalization of $(2n-1)$-dimensional Kepler problem]{An integrable generalization of $\boldsymbol{(2n-1)}$-dimensional\\ Kepler problem}\label{sec:5}

We present here an integrable Hamiltonian system which will be a natural generalization (preturbation) of regularized $(2n-1)$-dimensional Kepler problem discussed in Section~\ref{sec:4}.

Therefore, assuming for $z\in \mathbb{C}$ and $l \in \mathbb{Z}$ the convention
\begin{equation*}
z^l := \begin{cases}
z^l & \text{for } l\geq 0, \\
\bar{z}^{-l}& \text{for } l<0
\end{cases}
\end{equation*}
we define the following Hamiltonian
\begin{gather}
H = h_0 \big(|\eta_1|^2 , \ldots , |\eta_n|^2, |\xi_1|^2 , \ldots , |\xi_n|^2\big) + g_0 \big(|\eta_1|^2 , \ldots , |\eta_n|^2, |\xi_1|^2 , \ldots , |\xi_n|^2\big)\nonumber\\
\hphantom{H =}{}
\times \big(\eta_1^{k_1} \cdots \eta_n^{k_n}\xi_1^{l_1} \cdots \xi_n^{l_n} + \eta_1^{-k_1} \cdots \eta_n^{-k_n}\xi_1^{-l_1} \cdots \xi_n^{-l_n}\big),\label{65}
\end{gather}
on the symplectic manifold $\big(\mathbb{C}^{2n}, {\rm d}\gamma_{+-}\big)$, where $h_0$ and $g_0$ are arbitrary smooth functions of~$2n$ real variables and $k_1, \ldots k_n, l_1, \ldots , l_n \in \mathbb{Z}$. Let us note here that taking in (\ref{65}) $h_0 = I_{++}$ and $g_0 = 0$ we obtain $(2n-1)$-dimensional regularized Kepler Hamiltonian on $\mathcal{T}_{+-}^0/_\sim$. We see from~(\ref{65}) that $H$ is a radical generalization of $I_{++}$. Nevertheless, as we will show in the subsequent, the Hamiltonian system $\big(\mathcal{T}_{+-}^0/_\sim,H/_\sim\big)$ is still integrable in quadratures.

For this reason, according to \cite{KS}, we define, for $r=1,\ldots, 2n$, the functions
\begin{gather}\label{66}
I_r := \sum_{j=1}^n \rho_{r,j}|\eta_j|^2 - \sum_{j=1}^n \rho_{r, n+j} |\xi_j|^2, \\
\psi_r := \sum_{j=1}^n \kappa_{j,r}\phi_j + \sum_{j=1}^n \kappa_{n+j,r} \phi_{n+j},\nonumber
\end{gather}
where $\eta_j = |\eta_j|{\rm e}^{{\rm i}\phi_j}$, $\xi_j = |\xi_j|{\rm e}^{{\rm i}\phi_{n+j}}$. By definition the real $2n \times 2n$ matrix $[\rho_{r,s}]$ is invertible and the matrix $[\kappa_{r,s}]$ is its inverse. The functions $(I_1, \ldots , I_{2n}, \psi_1, \ldots , \psi_{2n})$ form a system of coordinates on the open subset
\begin{equation*}
\Omega^{2n} := \big\{ (\eta , \xi )\in \mathbb{C}^n \oplus \mathbb{C}^n\colon |\eta_1|\neq 0, \ldots , |\eta_n |\neq 0, |\xi_1|\neq 0, \ldots,|\xi_n|\neq 0 \big\}
\end{equation*}
of $\mathbb{C}^{2n}$.
They are a canonical coordinates for symplectic form ${\rm d}\gamma_{+-}$, i.e., their Poisson brackets satisfy
\begin{gather*}
\{I_r, I_s \} = 0, \qquad \{I_r, \psi_s \} =\delta_{rs}, \qquad \{\psi_r, \psi_s \} =0.
\end{gather*}
What is more, one easily checks that for $r=2, \ldots, 2n$ one has $\{H, I_r \} =0$ if and only if
\begin{equation}\label{57}
\sum_{j=1}^n (\rho_{r,j}l_j + \rho_{r, n+j}k_j) = \delta_{r1}.
\end{equation}
So, the Hamiltonian system on $\big(\Omega^{2n}, {\rm d}\gamma_{+-}\big)$ given by the Hamiltonian~(\ref{65}) is integrable and $H, I_2, \ldots , I_{2n-1}$ are its functionally independent integrals of motion in involution. Considering $(I_2, \ldots, I_{2n})$ as the components
\begin{equation}\label{mommapexm}
{\bf J}(\eta^+,\xi^+,\eta,\xi)= \left(\begin{matrix}
I_2(\eta^+,\xi^+,\eta,\xi)\\
\vdots \\
I_{2n}(\eta^+,\xi^+,\eta,\xi)
\end{matrix}\right)
\end{equation}
of the momentum map ${\bf J}\colon \Omega^{2n} \to \mathbb{R}^{2n-1}$, where one identifies $\mathbb{R}^{2n-1}$ with the dual space to the Lie algebra of $(2n-1)$-dimensional torus $\mathbb{T}^{2n-1} := \underbrace{U(1) \times \cdots \times U(1)}_{2n-1}$, we can apply Marsden--Weinstein reduction procedure to $\big(\Omega^{2n}, {\rm d}\gamma_{+-}, H\big)$. In this way we reduce the above Hamiltonian system to ${\bf J}^{-1}(c_2, \ldots, c_{2n})/ \mathbb{T}^{2n-1} \cong {}]a, b[{} \times \mathbb{S}^1$ with $\omega_{\rm red} = {\rm d}I_1 \wedge {\rm d}\psi_1 $ as a symplectic form, where $(I_1, \psi_1) \in {}]a,b[{} \times \mathbb{S}^1$, and the Hamiltonian~(\ref{65}) after the reduction to ${\bf J}^{-1}(c_2, \ldots, c_{2n})/ \mathbb{T}^{2n-1}$ assumes the following form
\begin{equation*}
H_{\rm red} = H_0 (I_1, c_2, \ldots, c_{2n}) + 2 \sqrt{G_0(I_1, c_2, \ldots, c_{2n})} \cos \psi_1,
\end{equation*}
where $H_0 (I_1, I_2, \ldots, I_{2n})$ and $G_0(I_1, I_2, \ldots, I_{2n})$ are defined as the superpositions of the functions
\begin{gather*}
h_0 \big(|\eta_1|^2 , \ldots , |\eta_n|^2, |\xi_1|^2 , \ldots , |\xi_n|^2\big)
\end{gather*}
 and
 \begin{gather*} \big(g_0 \big(|\eta_1|^2 , \ldots , |\eta_n|^2, |\xi_1|^2 , \ldots , |\xi_n|^2\big)\big)^2|\eta_1|^{2|k_1|} \cdots |\eta_n|^{2|k_n|}|\xi_1|^{2|l_1|} \cdots |\xi_n|^{2|l_n|}
 \end{gather*} with the map inverse to the map defined in~(\ref{66}). For the explicit expression for~$a$ and~$b$ see~\cite{KS}.

The Hamilton equations defined by $H_{\rm red}$ in the canonical coordinates $(I_1, \psi_1)$ have form
\begin{gather}\label{510}
\frac{{\rm d}I_1}{{\rm d}t} = 2 \sqrt{G_0(I_1, c_2, \ldots, c_{2n})} \sin \psi_1 , \\
\label{511}
\frac{{\rm d}\psi_1}{{\rm d}t} = \frac{\partial H_0}{\partial I_1} (I_1, c_2, \ldots, c_{2n}) + \frac{\partial G_0}{\partial I_1} (I_1, c_2, \ldots , c_{2n}) \cos \psi_1 .\nonumber
\end{gather}
From (\ref{510}) and $E:=H_{\rm red}(I_1(t), \psi_1 (t), c_2, \ldots, c_{2n}) = {\rm const}$, where~$E$ is the total energy of the system, we obtain
\begin{equation}\label{512}
\left(\frac{{\rm d}I_1}{{\rm d}t}\right)^2 = 4 G_0 (I_1, c_2, \ldots, c_{2n}) - (E- H_0(I_1, c_2, \ldots , c_{2n}))^2 .
\end{equation}
Separating variables in~(\ref{512}) we integrate it by quadratures. Next, using integrals of motion $I_2, \ldots, I_{2n}$, we integrate our initial system defined on~$\big(\mathbb{C}^{2n}, {\rm d}\gamma_{+-}\big)$ by the Hamiltonian~(\ref{65}). A~detailed description of this method of integration can be found in~\cite[Section~2]{KS}.

Now let us assume that the last two of integrals of motion $I_2, \ldots,I_{2n-1}, I_{2n}$ are given by
\begin{gather*}
I_{2n-1} :=I_{++} = \eta^+ \eta + \xi^+\xi, \qquad
I_{2n}:=I_{+-} = \eta^+\eta - \xi^+ \xi .
\end{gather*}
Hence, from (\ref{57}), we obtain the conditions
\begin{equation*}
k_1 + \dots + k_n =0 \qquad \text{and} \qquad l_1 + \dots + l_n =0
\end{equation*}
on the exponents $k_1, \ldots, k_n, l_1, \ldots, l_n \in \mathbb{Z}$, which guarantee integrability of the Hamiltonian system $\big(\mathbb{C}^{2n}, {\rm d}\gamma_{+-}, H\big)$. Because~$I_{+-}$ is one of the integrals of motion, we find that the reduced system $\big(\mathcal{T}_{+-}^0/_\sim , H/_\sim \big)$ is also integrable. So, using the symplectomorphism $\mathcal{C}_{\rm reg}\circ \mathcal{K}_{\rm reg}\colon \dot{\tilde{\mathcal{O}}}_{1,0} /_\sim \to \mathcal{T}_{+-}^0 /_\sim $, see diagram~(\ref{equivalence}), we obtain an integrable Hamiltonian system on $\dot{\tilde{\mathcal{O}}}_{1,0}/_\sim $ with Hamiltonian $(H/_\sim )\circ \mathcal{C}_{\rm reg}$.

In the particular case, if $k \in \{ k_1, \ldots , k_n\} $ and $ l\in \{l_1, \ldots , l_n\}$ then $ -k \in \{ k_1, \ldots , k_n
\} $ and $ -l\in \{l_1, \ldots , l_n\}$, the Hamiltonian (\ref{65}) depends on the matrix elements of $I_+$ and $I_-$ only. So, in this case we obtain from (\ref{516}) that the Hamiltonian $(H/_\sim )\circ \mathcal{C}_{\rm reg}$ could be defined as the reduction $\tilde{H}/_\sim $ to $\dot{\tilde{\mathcal{O}}}_{1,0}/_\sim$ of the Hamiltonian
\begin{gather}
\tilde{H} = h_0 \big(N^-_{11}, \ldots , N^-_{nn}, N^+_{11}, \ldots, N^+_{nn}\big)+ g_0 \big(N^-_{11}, \ldots , N^-_{nn}, N^+_{11}, \ldots, N^+_{nn}\big) \nonumber\\
\hphantom{\tilde{H} =}{} \times\big[\big(N^-_{i_1j_1}\big)^{k_{i_1}}\cdots \big(N^-_{i_rj_r}\big)^{k_{i_r}}\big(N^+_{a_1b_1}\big)^{l_{a_1}}\cdots \big(N^+_{a_sb_s}\big)^{l_{a_s}}+\text{h.c.}\big]\label{517}
\end{gather}
on $H(n)\times H(n)$, where $N^\pm_{kl}:=\frac{1}{2}( R_{kl}\pm M_{kl})$, $R$ and $M$ depend on $(Y,X)$ by (\ref{449}). The subsets of exponents $\{k_{i_1}, \ldots , k_{i_r}\} \subset \{k_1, \ldots , k_n\}$ and $\{l_{a_1}, \ldots , l_{a_s}\} \subset \{l_1, \ldots, l_n\}$ satisfy $k_{i_m}=-k_{j_m}>0$ for $m=1,2,\ldots , r$ and $l_{a_m}=-l_{b_m} >0$ for $m=1,2,\ldots, s$.

Ending this section, we write the Hamiltonian (\ref{517}) in the more explicit form for the case $n=2$. In this case the integrals of motion $M$ and $R$ can be written in terms of Pauli matrices
\begin{equation*}
M = M_0 E + \vec{M} \cdot \vec{\sigma} \qquad \text{and}\qquad R= R_0 E + \vec{R}\cdot \vec{\sigma} ,
\end{equation*}
where $\vec{M}$ and $\vec{R}$ are angular momentum and Runge--Lenz vector, respectively. Using the linear relation
\begin{equation*}
\left(\begin{matrix}
|\eta_1|^2\\
|\eta_2|^2\\
|\xi_1|^2\\
|\xi_2|^2
\end{matrix}\right) = \frac{1}{2} \left(\begin{matrix}
1&1&-1&-1\\
1& -1& -1&1 \\
1 & 1& 1& 1 \\
1& -1& 1& -1
\end{matrix}\right) \left(\begin{matrix}
R_0\\
R_3 \\
M_0\\
M_3 \end{matrix}\right)
\end{equation*}
and defining $M_+:= M_1 +{\rm i}M_2$ and $M_- := M_1 -{\rm i}M_2$ we write this Hamiltonian as follows
\begin{gather}
\tilde{H} = \tilde{h}_0 (R_0, R_3, M_0, M_3) + \tilde{g}_0 (R_0, R_3, M_0, M_3)\nonumber\\
\hphantom{\tilde{H} =}{}\times \big((R_{\sigma} - M_{\sigma})^k (R_{\sigma '} + M_{\sigma '})^l + (R_{-\sigma} - M_{-\sigma})^k (R_{-\sigma '} + M_{-\sigma '})^l\big),\label{520}
\end{gather}
where $\sigma, \sigma ' = +,-$, $k,l \in \mathbb{N}\cup \{0\}$ and $\tilde{h}_0$, $\tilde{g}_0$ are arbitrary smooth functions. Let us note that $R_0 = \frac{1}{2} I_0$. Note also that equation $M_0 = -\eta^+\eta + \xi^+\xi=0$ leads to the reduced system $\mathcal{T}_{+-}^0 /_\sim$.

In order to represent this Hamiltonian in the canonical coordinates $(\vec{y}, \vec{x})\in \mathbb{R}^3 \times \dot{\mathbb{R}}^3$, see~(\ref{456}), we note that
\begin{gather}
\label{521}\vec{M} = 2 \vec{x} \times \vec{y},\\
\label{522}\vec{R} = \big(1-{\vec{y}}^2\big)\vec{x} + 2 \vec{y}(\vec{x}\cdot\vec{y}).
\end{gather}
After substituting (\ref{521}), (\ref{522}) and $M_0=0$ and $R_0=||{\vec{x}}||\big(1+ (\vec{y})^2\big)$ into (\ref{520}) we reduce the Hamiltonian $\tilde{H}$ to the phase space $\big(\mathcal{S}(\Omega)\cong \mathbb{R}^3\times\dot{\mathbb{R}}^3, 2{\rm d}\vec{y}\wedge {\rm d} \vec{x}\big)$.

As it follows from the general method presented above, the Hamiltonian system on $\mathcal{T}^0_{+-}/ \sim $ described by the Hamiltonian (\ref{520}) for $n=2$, is integrable in quadratures, see equation~(\ref{512}). The third integral of motion complementary to $I_3= I_{++}$ and $I_4= I_{+-}$ is the following
\begin{equation*}
I_2 = \rho_{2,1}|\eta_1|^2 + \rho_{2,2}|\eta_2|^2 -\rho_{2,3}|\xi_1|^2-\rho_{2,4}|\xi_2|^2 ,
\end{equation*}
where the resonance condition
\begin{equation*}
(\rho_{2,1}- \rho_{2,2})l + (\rho_{2,3}-\rho_{2,4})k =0
\end{equation*}
is subjected to be fulfilled. In the Section~IV of the paper~\cite{efs}, where a perturbed Kepler problem (the hydrogen atom interacting with the constant electric and magnetic fields) is considered, the authors, using the normalization procedure, obtain an integrable approximation of the perturbed Kepler Hamiltonian investigated by them. See also~\cite{iwai3} for MIC-Kepler problem. This approximated system could be treated as a special subcase of~(\ref{520}), what follows from the fact that~(\ref{520}) is the general Hamiltonian, which has three Manley--Rowe type integrals of motion given by~(\ref{mommapexm}). The quantum version of the Hamiltonian system~(\ref{65}), as well as its integration by quantum reduction method, can be found in~\cite{KS}. Some methods of integration of a quantum perturbed Kepler system can be found in~\cite{efs}. All these questions for the integrable generalized $(2n-1)$-Kepler problem defined by the Hamiltonian~(\ref{65}) will be a subject of the next paper.

\subsection*{Acknowledgements}

Author would like to express his sincere gratitude for all the anonymous referees for their comments and remarks which improved the paper and made it more readable.

\pdfbookmark[1]{References}{ref}
\LastPageEnding

\end{document}